# Dielectric control of reverse intersystem crossing in thermally-activated delayed fluorescence emitters

*Alexander J. Gillett[1*#], Anton Pershin[2,3#], Raj Pandya[1], Sascha Feldmann[1], Alexander J. Sneyd[1], Antonios M. Alvertis[1], Emrys W. Evans[1,4], Tudor H. Thomas[1], Lin-Song Cui[1], Bluebell H. Drummond[1], Gregory D. Scholes[5], Yoann Olivier[6], Akshay Rao[1], Richard H. Friend[1] and David Beljonne[2*].*

[1]Cavendish Laboratory, University of Cambridge, JJ Thomson Avenue, Cambridge, CB3 0HE, UK.

[2]Laboratory for Chemistry of Novel Materials, Université de Mons, Place du Parc 20, 7000 Mons, Belgium.

[3]Wigner Research Centre for Physics, PO Box 49, H-1525, Budapest, Hungary.

[4]Department of Chemistry, Swansea University, Singleton Park, Swansea, SA2 8PP, UK.

[5]Department of Chemistry, Princeton University, Princeton, New Jersey 08544, United States.

[6]Unité de Chimie Physique Théorique et Structurale & Laboratoire de Physique du Solide, Namur Institute of Structured Matter, Université de Namur, Rue de Bruxelles, 61, 5000 Namur, Belgium.

[*]Corresponding authors: David Beljonne: E-mail: david.beljonne@umons.ac.be; Alexander J. Gillett: E-mail: ajg216@cam.ac.uk.

[#]These authors contributed equally.

**Thermally-activated delayed fluorescence (TADF) enables organic semiconductors with charge transfer (CT)-type excitons to convert dark triplet states into bright singlets *via* a reverse intersystem crossing (rISC) process. Here, we consider the role of the dielectric environment in a range of TADF materials with varying changes in dipole moment upon optical excitation. In a dipolar reference emitter, TXO-TPA, environmental reorganisation after excitation in both solution and doped films triggers the formation of the full CT product state. This lowers the singlet excitation energy by 0.3 eV and minimises the singlet-triplet energy gap ($\Delta E_{ST}$). Using impulsive Raman measurements, we observe the emergence of two (reactant-inactive) modes at 412 and 813 cm$^{-1}$ as a vibrational fingerprint of the CT product. In contrast, the dielectric environment plays a smaller role in the electronic excitations of a less dipolar material, 4CzIPN. Quantum-chemical calculations corroborate the appearance of these new product modes in TXO-TPA and show that the dynamic environment fluctuations are large compared to $\Delta E_{ST}$. The analysis of the energy-time trajectories and the corresponding free energy functions reveals that the dielectric environment significantly reduces the activation energy for rISC, thus increasing the rISC rate by up to three orders of magnitude when compared to a vacuum environment.**

Organic light emitting diodes (OLEDs) based on materials that exhibit thermally activated delayed fluorescence (TADF) have gained significant attention for their ability to utilize dark triplet excitons for light emission *via* a thermally-assisted reverse intersystem crossing (rISC) process[1–3]. As a result, the internal quantum efficiencies of TADF OLEDs can now approach 100%, with virtually all injected charge carriers exploited for light emission[1–6]. The original central design criteria for a TADF emitter was to ensure the spatial separation of the hole and electron wavefunctions (usually approximated by the highest occupied molecular orbital (HOMO) and lowest unoccupied molecular orbital (LUMO), respectively) by utilising separate electron donor (D) and acceptor (A) moieties; this results in singlet and triplet states with significant intramolecular charge transfer (CT) character[1]. The consequence of decoupling of the HOMO and LUMO is a small overlap integral for the electron wavefunctions, which minimizes the electron exchange energy (J) and therefore the singlet-triplet energy gap ($\Delta E_{ST}$ ~2J)[7]. However, a rISC process between triplet and singlet states that both have the same CT character ($^3$CT and $^1$CT) is formally forbidden; according to El-Sayed's rule, the spin-orbit coupling (SOC) matrix elements in this case are vanishingly small (<0.1 cm$^{-1}$)[8,9]. Thus, it has been widely-accepted that the presence of energetically-close triplet states with different natures with respect to the CT states is critical for efficient TADF[10–13]. These triplet states are local excitons ($^3$LE) located on either the D or A. Multiple different roles of the $^3$LE states in the rISC process have been proposed, including vibronic mixing of the $^3$LE with the $^3$CT and the $^3$LE acting as an intermediate between the $^1$CT and $^3$CT states, through which SOC-mediated rISC is formally allowed[11–14].

However, despite the substantial effort to understand the mechanism of rISC in TADF emitters, relatively little attention has been given to the role of the surrounding dielectric environment in this process[9,15–17]. In OLED devices, the TADF dopant is commonly dispersed in a wide band-gap host at low weight percentages (1-10 wt%) to suppress concentration-quenching effects that reduce the efficiency of luminescence[18]. Thus, whilst some aggregation of chromophores may occur[19], the TADF dopant will experience strong host-guest interactions that are likely to have a significant impact on their photophysical properties. Indeed, due to the CT nature of their excited states, TADF materials are liable to exhibit a very large change in their dipole moment upon optical or electrical excitation, which can be on the order of 20 D or larger[16,20]. This change in dipole moment will induce a reorientation of the local dielectric environment to better stabilize the new electrostatic configuration of the chromophore[21,22]. Such effects are not limited to solution environments where the solvent can freely rotate[23–25], but have also been observed in films comprised of TADF emitters dispersed in both polymer and small molecule hosts[16,26,27]. These environmental reorganization effects can have a significant impact on the properties of the molecule, with large changes in the optical band gap, the coupling between electronic states and even in the relative energetic ordering of the CT and $^3$LE states possible[16,26].

**Results**

In this study, we investigate a range of high-performance TADF materials that represent popular structural motifs employed by the field, including linear D-A (TXO-TPA and DACT-II)[28,29], D-A-D (2CzPN)[1] and multi-D systems (4CzIPN and 5Cz-TRZ)[1,23]. The chemical structures of the emitters are displayed in Fig. 1a. In addition, the structures of the three wide-gap host materials chosen to represent solution (toluene), small molecule (UGH2) and polymer (polystyrene: PS) host environments are also included (Fig. 1b). All host materials are chosen to possess both similar structural motifs and comparable dielectric constants of between 2.4-2.8 in order to restrict the environmental variables to, as far as possible, only the nature of the host[30–32]. This is important, as previous theoretical studies have predicted that the host dielectric constant can have a large effect on the nature of the singlet and triplet excited states and their energy separation[33]. To begin, we focus on TXO-TPA as a model system due to its relatively simple D-A structural motif and prominence as a high-performance material[28,34]. We note that the steady-state photoluminescence (PL) spectra of TXO-TPA in both polymer and small-molecule hosts are very similar to the toluene solution, with only small shifts in the PL maxima of ±10 nm (Fig. S1). This confirms that the energy of the relaxed TXO-TPA $^1$CT state is similar in all environments, enabling a comparative study on the role of the host dynamics in the (r)ISC processes of this TADF emitter to be undertaken.

We have performed pump-probe transient absorption (TA) spectroscopy on a solution of TXO-TPA in an oxygen-free toluene (Fig. 2a). Here, TXO-TPA was excited with a 400 nm (100 fs) pulse to populate the $^1$CT state. Beginning 200-300 fs after excitation, we observe two photo-induced absorption (PIA) features: a strong band in the near-infrared (NIR) peaked at 975 nm and a broad shoulder at 625 nm that extends into the visible region. Subsequently, a rapid spectral evolution occurs, with the NIR PIA blue-shifting from 975 nm to 925 nm and the PIA shoulder at 625 nm reducing in intensity to give two new peaks at 520 nm and 770 nm. We have taken kinetics at wavelengths associated with these key features to examine their temporal evolution: one from the low (1000-1025 nm) and high (920-940 nm) energy edges of the NIR PIA and another from the shoulder (620-650 nm) of the PIA (Fig. 2b). In all kinetic traces, the excited state evolutions are largely complete by 20 ps. We notice similar spectral shifts in the PIA features on picosecond timescales for the other four TADF materials in toluene solutions (Figs. S2-S5), though these are noticeably reduced in 4CzIPN and 5Cz-TRZ. Calculations of the change in dipole moment in the TADF emitters upon excitation from the ground state to the $^1$CT reveal a clear correlation with the magnitude of the spectral shifts (Fig.

1c), with the largest reported for the most dipolar systems (TXO-TPA and DACT-II). Thus, we conclude that these PIA shifts result from the local environment reorganising in response to the new CT electrostatic configuration of the chromophore[21,22], which then induces further delayed changes in the molecular and electronic structure of the emitter (*vide infra*). Indeed, our observation of this behaviour in a range of materials with diverse structural features suggests that it may be an innate part of the TADF process.

To supplement the TA measurements on the TADF emitters in a toluene solution, we have also investigated the emissive behaviour on sub-nanosecond timescales with transient grating PL (TGPL) spectroscopy. We focus here on two TADF materials with contrasting PIA peak shift sizes: TXO-TPA and 4CzIPN. For consistency with the TA, both materials were excited at 400 nm. In TXO-TPA (Figs. 2c and d), the PL maxima is 510 nm at 200-300 fs after excitation; significantly higher in energy than the steady-state PL, which peaks at 580 nm (Fig. S1). We then observe a red shift of the PL to 570 nm over identical timescales to the PIA shifts. This picosecond Stokes shift of ~0.3 eV in the PL position confirms that the PIA shifts in the TA result from the environment-mediated stabilization of the $^1$CT electronic configuration. We note that these timescales are consistent with previous observations of solvent-mediated relaxation in materials with CT-type excited states[25,35–37]. We also observe a large increase in the PL intensity towards 1 ns for the band centred at 570 nm. We attribute this to the onset of the emission from the fully relaxed $^1$CT, in-line with a typical fluorescence rate constant of ~$10^7$ s$^{-1}$ in TADF materials[38]. This delayed rise in the relaxed $^1$CT PL, consistent with the small radiative rate expected for a CT-type excitation, also implies that the state responsible for the blue-shifted emission on sub-picosecond timescales must have a signficantly larger oscillator strength and an increased electron-hole overlap. Therefore, we propose that the dielectric environment drives the formation of a quasi-pure CT excitation in TXO-TPA from an initial state with only partial CT character. In contrast, there is no signficant PL detected from 4CzIPN on early timescales <10 ps (Fig. S6); instead, the luminescence grows in towards 1 ns. This indicates that after optical excitation, 4CzIPN directly forms a low oscillator strength $^1$CT state with a small electron-hole overlap. In addition, we observe no spectral shifts in the 4CzIPN PL on sub-nanosecond timescales, which already matches the steady-state PL (Fig. S7). This also signifies that there is minimal delayed electronic or structural evolution induced by the environment in 4CzIPN following excitation to the $^1$CT state. These observations are consistent with the limited reorganisation of the surrounding dielectric environment that is expected in response to the small change in the dipole moment of 4CzIPN upon excitation into the $^1$CT state.

To explore whether the dielectric environment can also influence the electronic processes in the solid state, we have performed identical oxygen-free measurements on TXO-TPA dispersed in PS at 5 wt% and UGH2 at 10 wt%. In both solid-state hosts, we observe similar behaviour to that of TXO-TPA in toluene, with blue-shifts from the initial NIR PIA peak position and concomitant red shifts of the PL peak from the starting position at ~510 nm (Figs. S8-S11). However, we note that the spectral shifts occur more slowly in the films. Additionally, by 1.5 ns, the PL peak does not reach the wavelength seen in the steady-state measurements of 570 nm and 590 nm for the PS and UGH2 films, repectively. Therefore, we have also explored the nanosecond trPL of these two films (Figs. S12-S13). Here, we see a continued PL red-shift over nanosecond timescales that is not detected for TXO-TPA in toluene (Fig. S14). The red-shift to the steady-state PL maxima completes by ~100 ns in the PS host and ~20 ns in UGH2. The presence of a nanosecond PL red-shift is in-line with previous reports of TADF emitters in both polymer and small molecule hosts, attributed to the slower reorganisation in the solid-state environment[20,22]. We note that it is also possible for slow downhill energy migration within the ensemble of different chromophore molecular configurations to contribute to the later time red-shift in the solid films[39], though both intramolecular reorganisation and energy migration will explore a similar region of conformational space and ultimately form the same lowest energy $^1$CT state. However, our observation of the picosecond stabilisation component of the $^1$CT of TXO-TPA confirms that solid-state dielectric environments are indeed capable of a rapid reorganisation in response to a change in the electronic configuration of the TADF emitter[16,26,27].

Having fully characterized the ultrafast TA and TGPL of TXO-TPA and 4CzIPN, we next probe the nature of the pre- and post-environment reorganization state with impulsive vibrational spectroscopy (IVS)[40]. 4CzIPN was excited with a 450 nm (200 fs) pump pulse to populate the $^1$CT electronic state directly. Owing to its weaker $^1$CT absorption band, TXO-TPA was instead excited with a 343 nm (270 fs) pump pulse. Importantly, the observed behaviour of TXO-TPA excited at 343 nm is equivalent to excitation at 400 nm, with all spectral evolutions again completed by 20 ps (Fig. S15). Subsequently, a second ultrafast (<10 fs; shorter than the period of most vibrational modes) 'push' pulse, resonant with the $^1$CT PIA, was used to generate vibrational coherences in the molecular normal modes of the $^1$CT potential energy surface. To resolve the excited state vibrational modes from the ground state modes, IVS requires a large fraction (~1-10%) of the molecules under investigation to be excited. This necessitates the use of very high excitation fluences, which can lead to rapid material degradation in solid films. Thus, we focus here on the toluene solutions. However, our observations of similar ultrafast stabilisation of the $^1$CT in both solution and solid-state environments for TXO-TPA indicates that our findings in solution will also be applicable to the solid-state. In TXO-TPA, the kinetic traces averaged over the 780-940 nm spectral range of the $^1$CT PIA as a function of push to probe delay show strong temporal oscillations (Figs.

S16-S18). Fourier transforms of these oscillations at varying pump-push delays of 500 fs, 3 ps and 10 ps, are shown in Fig. 3. This allows us to probe before, during and after the environment-mediated stabilization of the $^1$CT. Residual toluene solvent modes are marked with a black asterisk (the IVS spectrum of toluene is provided in Fig. S19). To enable the identification of modes characteristic of the $^1$CT excited state, the off-resonant Raman of a neat TXO-TPA film is also presented with the IVS plots. In TXO-TPA, there are high activity low-frequency components present (<100 cm$^{-1}$) at all push times, likely associated with torsional modes modulating the D:A dihedral angle[41]. However, it is not possible to resolve the individual modes present in this region. The key observation is the gradual appearance of two new intense modes at 412 cm$^{-1}$ and 813 cm$^{-1}$ (denoted by red asterisks) at 3 and 10 ps that are not present 500 fs after excitation by the pump pulse. These modes can be considered a fingerprint of the quasi-pure CT excitation in TXO-TPA and will be explored in more detail with the quantum-chemical calculations. In contrast, we do not see the formation of new vibrational modes in the IVS of 4CzIPN, taken over the $^1$CT PIA between 800-865 nm: only limited changes in the mode intensities (Fig. S20). The more subtle evolution of the vibrational modes is consistent with the weaker effect of the dielectric environment reorganisation on the molecular and electronic structure of 4CzIPN.

We now discuss the results of the combined quantum mechanics/molecular mechanics (QM/MM) adiabatic dynamics simulations of TXO-TPA in a dilute toluene solution. We have focussed here on TXO-TPA due its relative structural simplicity and large change in dipole moment upon excitation, but we consider our findings generally applicable to TADF materials that also exhibit large changes in their dipole moments. Furthermore, the observation of picosecond timescale stabilisation in dilute dispersions of TXO-TPA in solid matrices confirms that our conclusions are also of relevance to the processes occurring in operational OLED devices. In the Franck-Condon region, the (vertical) excitation energy to the $^1$CT exceeds that of the lowest T$_1$ ($^3$CT-like) and T$_2$ ($^3$LE-like) states by 0.4 and 0.1 eV, respectively (Table S1). In the time evolution of the $^1$CT excitation energy (see one representative trajectory in Fig. 4a), we observe a complex oscillatory evolution over the whole 10 ps trajectory explored. Ultimately, we find that these nuclear motions dump ~0.3 eV (from ~3.1 eV at 0 ps to ~2.8 eV at 10 ps) into the environment, consistent with the experimental observations of a ~0.3 eV Stokes-shift in the PL maximum of TXO-TPA over these timescales. Indeed, we notice a shift in the distributions of the $^1$CT energy to lower values when moving from the early (0-3 ps) to late (3-10 ps) timescales of the explored trajectory (Fig. 4b).

To explore the role of the solvent in this slow fluctuation, we have computed the electrostatic potential generated by the toluene molecules on TXO-TPA over the simulation trajectory (Fig. S27). The results, imaged in Fig. 4c for several snapshots extracted along the *ab initio* MD trajectory, demonstrate how the toluene molecules reorient in response to the dipolar field generated upon photoexcitation of the TXO-TPA solute to the $^1$CT state. It appears that the solvent orientational relaxation response drives the formation of a highly polarized electrostatic environment by 2.5 ps; this is in clear contrast to the largely unstructured electrostatic pattern in the ground-state solvent configuration at 0.3 ps. Importantly, the collective orientation polarization of the weakly polar toluene molecules is strong enough to lower the TXO-TPA $^1$CT excitation energy by ~0.3 eV after 10 ps. Through analysing the temporal evolution of the electron-hole overlap and the transferred charge in the excited state wavefunction (Fig. S28), we conclude that the decrease in $^1$CT energy is associated with an increased CT character of the lowest singlet excited (S$_1$) state. This is reminiscent of twisted intramolecular CT (TICT), where electron transfer from the D to A involves a change in conformation (rotation around the D:A single bond), and also shows a strong dependence of the local environment[42]. This is seen in the time evolution of the dihedral angle between the TPA D and the TXO A moieties that rises from ~50° to ~90° after 4-5 ps (Fig. 4d). Importantly, this effect is synergistically driven by, but also in response to, solvent polarization effects.

One consequence of the $^1$CT state acquiring an increasing CT character is a sharp drop in the electron-hole interaction and hence a fall in $\Delta E_{ST}$ from the initial ~0.4 eV in the ground-state equilibrium geometry down to values as small as ~0.1 eV (Fig. 4e). We also note that: (i) as expected due to their negligible dipole moment[43], the $^3$LE states are only weakly affected by the environment reorganization, remaining at a near-constant energy (as a result, the $^1$CT and $^3$CT sweep below the $^3$LE, see Fig. S29); (ii) the initial solvent reorganization, completed after a few picoseconds, prompts large and sustained fluctuations at later times in the energy and nature of the low-lying electronic excitations. For example, variations in the $^1$CT energy and wavefunction form (as measured from the electron-hole overlap, $\phi_S$) between 3 and 10 ps after excitation result in large standard deviations of these values: 0.29 eV and 7x10$^{-3}$, respectively. The amplitude of the fluctuations is connected to the large reorganization energy ($\lambda$) of the CT state versus the ground state (owing to a fluctuation-dissipation relation) and is expected to be approximately equal to $(\lambda kT)^{1/2}$, where $k$ is the Boltzmann constant and $T$ is temperature. We note that these fluctuations are explicitly associated with a combination of solvent and solute nuclear motions. Thus, it is clear that the interplay between molecular reorganization and solvent response (inner- and outer-sphere reorganization) is critical in determining the $\Delta E_{ST}$ between $^1$CT and both $^3$CT and $^3$LE in TADF materials, which, in turn, is a key factor controlling the (r)ISC processes[43].

To examine the intramolecular reorganization, we have decomposed the predicted spectral dynamics for the $^1$CT excitation energies onto the vibrational modes of TXO-TPA as it explores the $^1$CT MD trajectory; all vibrational modes are displayed in Fig. 5a, with the molecular localisation of the modes shown in Fig. S30. This provides an instantaneous picture for the modes that contribute the most (those with the highest intensity) to the overall geometric distortion at any given time after excitation. To further visualise this, we have computed the standard deviations of the mode intensities along specific time intervals; this highlights changes in the relative activity of these modes with time (Fig. 5b). By comparing the system at 'short' (<3 ps) and 'long' (from 3 to 10 ps) times after excitation, we examine the reorganization process. We find that the low-frequency modes (<200 cm$^{-1}$) are present across the full simulation timescale, except for the first few hundred femtoseconds where these modes have not yet fully developed due to their relatively long oscillation periods (>0.17 ps) and the classical model for nuclei motion in our simulations. These low-frequency modes primarily involve changes in the D:A torsion angles that modulate the through-bond D:A interactions and act mostly as spectator modes, namely they are active in both the (Franck-Condon) reactant state and the (quasi-pure CT) product state, see Fig S31. The interplay of solvent relaxation and geometry-change activates two clusters of molecular vibrational modes at frequencies around ~400 and ~700-800 cm$^{-1}$, in good agreement with the IVS data. The two strongest modes at 417 and 712 cm$^{-1}$ correspond to vibrations involving both the TPA D and the TXO A moieties (thus extending over the entire molecule, see Fig. S32), with large contributions on the phenyl ring bridging the D and A units. These vibrational modes are absent in the reactant state (<5 ps) and their emergence indicates formation of the TICT product. Similar reactive modes have been reported recently for an intermolecular electron transfer reaction in solution by Rafiq *et al.*[44].

The picture that these calculations suggest is that multiple spectator modes (including dihedrals) couple to the TXO-TPA molecule in the reactant, Frank-Condon, state (showing partial CT). As the solvent polarizes around the solute, the electronic $S_1$ state acquires increasing CT character or, in other words, electron transfer occurs adiabatically from TPA to TXO, ultimately turning the system into a TICT product. The formation of the product switches on vibration modes associated with the reaction, mostly extended vibrations peaking around ~400 and ~700-800 cm$^{-1}$ for TXO-TPA in toluene, that should be regarded as fingerprints for the formation of the quasi-pure CT excited state. Relaxation proceeding from this novel non-equilibrium configuration then results in the irreversible formation of the $^1$CT product.

The effect of dielectric environment reorganization on the triplet manifold is difficult to access directly from experiment. To gain insight into this phase of the dynamics, we have repeated the QM/MM/MD calculations for the lowest adiabatic triplet state; in Fig. 6 we report the results of one 10 ps long MD trajectory. The situation is more complex in the triplet manifold, as the fluctuations in the environment and molecular vibrations now act in tandem to swap the nature of the lowest energy triplet excitation. This is visualized in Fig. 6a, where the time evolution of the electron/hole overlap in the lowest two triplet states exhibits sudden jumps from values as high as ~0.8 (indicative of a dominant $^3$LE character) to values as low as ~0.2 (corresponding to mostly $^3$CT excitations) for times <5 ps. After 5ps, the nature of the lowest energy triplet state swaps less frequently, often (but not always) possessing $^3$CT character. We note that the lowest energy (and therefore populated) triplet state has a low electron-hole overlap during the ~5-6.5 ps time range of our MD simulations, indicating $^3$CT character. Correspondingly, a polarized electrostatic pattern develops in the environment during this metastable time window (Fig. S33).

One consequence of the development of a polarized environment that acts to stabilize the $^3$CT is that the $^1$CT, with a similar nature, is also stabilized. However, the energy of the $^3$LE is affected to a much lesser degree as it possess a much smaller molecular dipole moment[22]. As a result, we note that the $S_1$, $T_1$ and $T_2$ simultaneously become confined within a narrow energy spacing in the 5-6.5 ps time window of this specific trajectory (Fig. 6b). Critically, this provides the opportunity for population transfer to take place between the singlet and triplet manifolds, which has been shown to occur at the crossing seams of the potential energy surfaces of the states involved[45]. As expected, the SOC between the $S_1$-$T_1$ and $S_1$-$T_2$ states takes a slightly lower average value during this time window (Fig. S34), though this trend is blurred by the large thermal fluctuations in SOC that are present throughout the simulation trajectory. Regardless, the SOC between $S_1$-$T_1$ and $S_1$-$T_2$ in TXO-TPA remains at ~0.1-0.3 and ~0.5-1.0 cm$^{-1}$, respectively; large enough to drive rISC[23], provided the singlet and triplet states remain degenerate for long enough. Thus, we propose that the interplay of solvent and intramolecular reorganization, seen in our trajectories as stochastic environmental fluctuations, give rise to equilibrium conditions that are favourable for rISC to take place (as occurs during the 5-6.5 ps time window of the trajectory reported in Fig. 6); this provides evidence for a rISC process in dipolar TADF emitters that has a strong contribution form the dielectric environment, comparable to previous observations that the rISC process can be enhanced by intramolecular reorganization of the TADF molecule[9].

We have also performed the same normal-mode analysis as in the singlet case but now using the lowest triplet as the target state. The results shown in Fig. 6c parallel those obtained for the $^1$CT; in particular, the bands at ~400 and ~700-800 cm$^{-1}$, associated with the formation of the quasi-pure CT excited state, become active at >5 ps. Crucially, this is consistent with the timescales over which the polarized environment that stabilizes the $^3$CT state develops. Therefore, once the solvent polarization stabilizes the $^3$CT for a sufficiently long time, it turns on the vibrational modes that are able to couple the two spin manifolds, driving triplet-to-singlet conversion[46].

To demonstrate the contribution from the dielectric environment to the (r)ISC processes, we have transformed 25 ps of singlet and 30 ps of triplet trajectories for both an explicit toluene solvent and a vacuum environment (trajectories in Fig. S35) into reactant and product free energy functions, $\Delta g_\alpha(X)$, using the $S_1$-$T_1$ energy gap as generalized microscopic reaction coordinate (Figs. 7a and 7b)[47]. For the reactant, this is given by:

$$\Delta g_\alpha(X) = -kT \ln[P(X)_\alpha] \qquad (1)$$

where $(X)_\alpha = E_\beta - E_\alpha$, with $\alpha$ denoting the reactant and $\beta$ the product; and $P(X)_\alpha$ is the probability that the system will have a given value of $X$ along the trajectories propagated on state $\alpha$. Thus, $\alpha \equiv S_1$ and $\beta \equiv T_1$ for the singlet trajectories; a similar free energy function can be obtained for the product, ($\alpha \equiv T_1$ and $\beta \equiv S_1$) from the triplet trajectories, now adding the reaction free energy, $\Delta G^0$, to Eq. 1. Here, it is important to note that the free energy function for the triplet is that of the lowest adiabatic $T_1$ state, which involves a dynamically evolving mixture of $^3$CT and $^3$LE excitations (Fig. 6a). In effect, the functions of $\Delta g_\alpha(X)$ are the microscopic equivalents of the Marcus parabola, where the minima represent the most frequently encountered $S_1$-$T_1$ energy gap in the trajectories; a free energy gain is required to move away from the minima, as expected for less commonly occurring $S_1$-$T_1$ gaps. Differently to Aizawa *et al.*, we note the appearance of a crossing seam between $S_1$ and $T_1$ (at $X = 0$, by definition of the free energy functions) as these involve (slightly) different electronic configurations with sufficient SOC to drive rISC. From the free energy curves, we extract a mean reorganisation energy (*λ*) of 210 meV and $\Delta G^0$ =95 meV for TXO-TPA in the explicit toluene solvent; these values correspond to an activation energy ($E_A$) of ~15 meV for ISC and ~110 meV for rISC. In clear contrast, we obtain $\lambda$ =750 meV and $\Delta G^0$ =140 meV for TXO-TPA in vacuum; this translates into a significantly higher $E_A$ of ~124 meV for ISC and ~264 meV for rISC.

To examine the impact of the dielectric environment on the rate of ISC ($k_{ISC}$) and rISC ($k_{rISC}$), we input the values obtained from our free energy functions in Figs. 7a and 7b into a classical Marcus-type non-adiabatic expression[45], taking a conservative estimate for the SOC of 0.5 cm$^{-1}$ from our simulations (Fig. S34). From this, we obtain $k_{ISC}$ =7.6×10$^7$ s$^{-1}$ and $k_{rISC}$ =1.8×10$^6$ s$^{-1}$ for the explicit toluene and $k_{ISC}$ =6.0×10$^5$ s$^{-1}$ and $k_{rISC}$ =2.5×10$^3$ s$^{-1}$ for the vacuum environment. We note that $k_{ISC}$ obtained when including the dielectric contribution is in excellent agreement with that determined experimentally for TXO-TPA in toluene ($k_{ISC}$ =5.7×10$^7$ s$^{-1}$), with reasonable agreement found for the experimental $k_{rISC}$ =2.3×10$^5$ s$^{-1}$ (see SI for details[48]). However, the consistency between the calculated and experimental $k_{rISC}$ is further improved, with little effect on $k_{ISC}$, by instead using a semi-classical Marcus-Levich-Jortner expression with one vibrational mode treated quantum mechanically (either the 400 or 800 cm$^{-1}$ modes identified for TXO-TPA in our work) and an internal contribution to the reorganisation energy of ~160-190 meV (the total reorganisation energy is kept constant)[49], see Fig. S36. Thus, it clear that even a weakly polar dielectric environment, such as toluene, can make a significant contribution to the rISC (and ISC) process in a dipolar TADF emitter.

**Discussion**

In this work, we have demonstrated the crucial role the dielectric environment plays in controlling the electronic properties of TADF emitters. We find experimental evidence, supported by quantum-chemical calculations, that the environment reorganizes over picosecond timescales to stabilize the dipolar CT-type electronic excitations created in TADF materials after optical (or indeed electrical) excitation by up to ~0.3 eV. This stabilization results from the environment-mediated formation of a $^1$CT excited state with increased CT character, which also simultaneously minimizes $\Delta E_{ST}$. Furthermore, the interplay of environment and intramolecular reorganization present in both the singlet and triplet manifolds dynamically varies the electronic properties of the TADF material that control the (r)ISC processes, including the $\Delta E_{ST}$, SOC between $S_1$-$T_1$ and $S_1$-$T_2$, and the nature of the lowest-lying triplet states. As a result, even a weakly polar environment such as a toluene solution can significantly reduce the activation energy for rISC in a dipolar emitter, effectively increasing $k_{rISC}$ by up to three orders of magnitude when compared to the isolated molecule. While this dynamic evolution is likely slower compared to solution, we expect the TADF molecules will explore a similar region of configurational space in solid films, with a possible contribution from energy transfer between chromophores with (partly) locked molecular conformations and environments. Furthermore, due to our observations of significant dielectric influence on the electronic properties of a diverse range of TADF emitters (both in solution and the solid state), we propose that our findings about the role of the dielectric environment in the rISC process of TADF materials will be generally applicable to those that exhibit a large change in their dipole moment upon excitation. This represents many materials with D-A or D-A-D structural motifs.

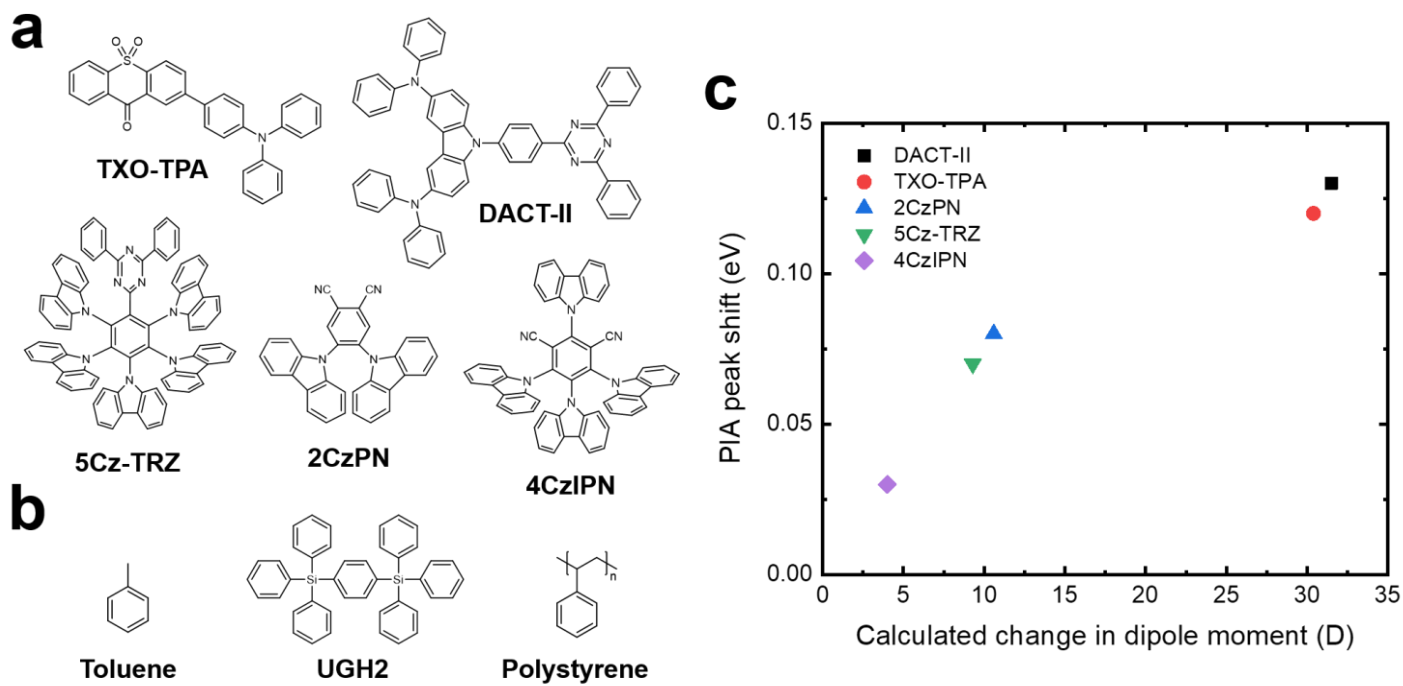

**Figure 1: (a)** The chemical structures of the five TADF emitters investigated in this study. The emitters were chosen to provide a representative selection of the popular structural motifs employed in high-performance TADF materials. This includes D-A (TXO-TPA), D-A-D (2CzPN and DACT-II) and multi-D (4CzIPN and 5Cz-TRZ). **(b)** The chemical structures of the host materials investigated in this study, chosen to represent solution (toluene), small molecule (UGH-2) and polymer (polystyrene). **(c)** The measured shift in the $^1$CT photo-induced absorption (PIA) peak between 0.3 and 100 ps (in eV) as a function of the calculated change in the dipole moment of the emitter upon excitation from the ground state to the $^1$CT state. In 2CzPN where more than one $^1$CT PIA is present, the PIA with the largest energy shift is presented.

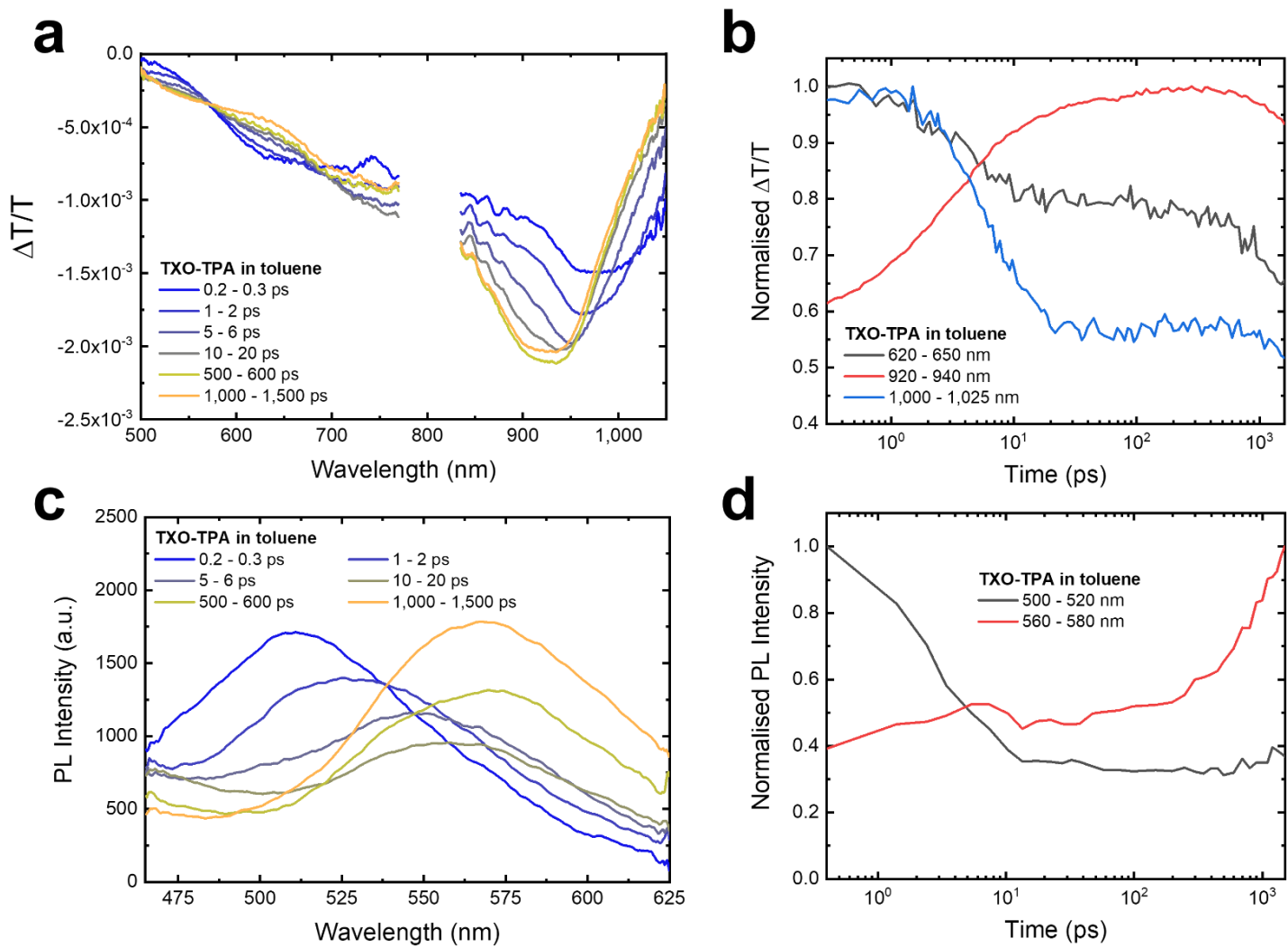

**Figure 2: (a)** The TA spectra of TXO-TPA in a toluene solution, excited at 400 nm with a fluence of 15.6 μJ/cm$^2$. **(b)** The TA kinetics of TXO-TPA in toluene, taken from the SE region (620-650 nm), the high energy edge of the $^1$CT PIA (920-940 nm) and the low energy edge of the $^1$CT PIA (1000-1025 nm). **(c)** The TGPL spectra of TXO-TPA in a toluene solution, excited at 400 nm with a fluence of 50.9 μJ/cm$^2$. **(d)** The kinetics of the TGPL, taken at the high energy (500-520 nm) and low energy (560-580 nm) edges of the TXO-TPA PL.

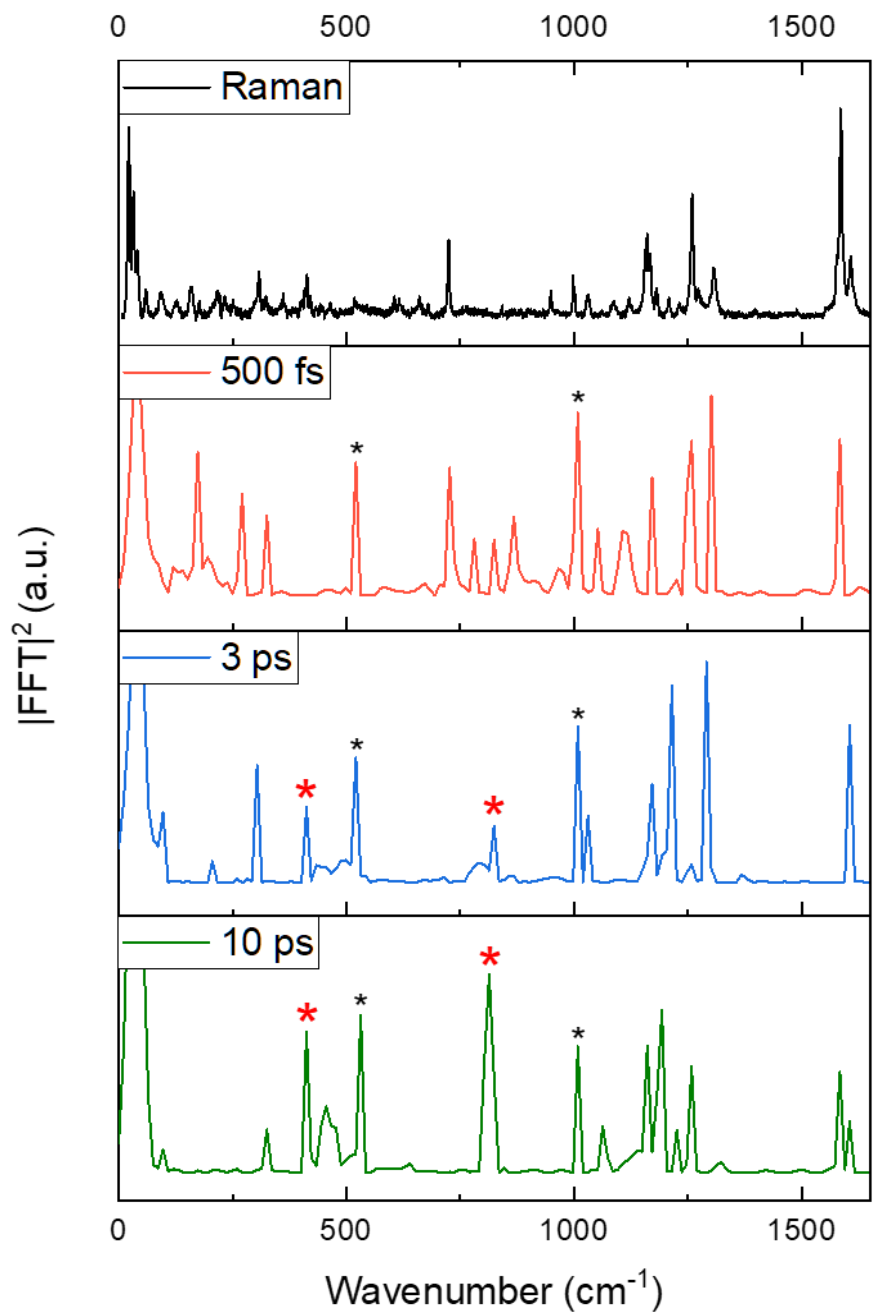

**Figure 3:** The top panel displays the steady state off-resonant (633 nm) Raman spectra of a neat TXO-TPA film. This can be compared to the IVS spectra of TXO-TPA in toluene taken at 0.5, 3 and 10 ps after excitation to isolate the modes that are associated with the $^1$CT excited state of TXO-TPA; the key modes associated with the formation of the quasi-pure CT state at 412 and 813 cm$^{-1}$ are denoted by red asterisks (toluene solvent modes are marked by black asterisks). The IVS was obtained by first exciting TXO-TPA at 343 nm with a 270 fs pulse (fluence 0.75 mJ/cm$^2$), before a second pulse centered at 790 nm with a 9.8 fs duration was used (1.1 mJ/cm$^2$) to induce coherent oscillations in the excited state of TXO-TPA.

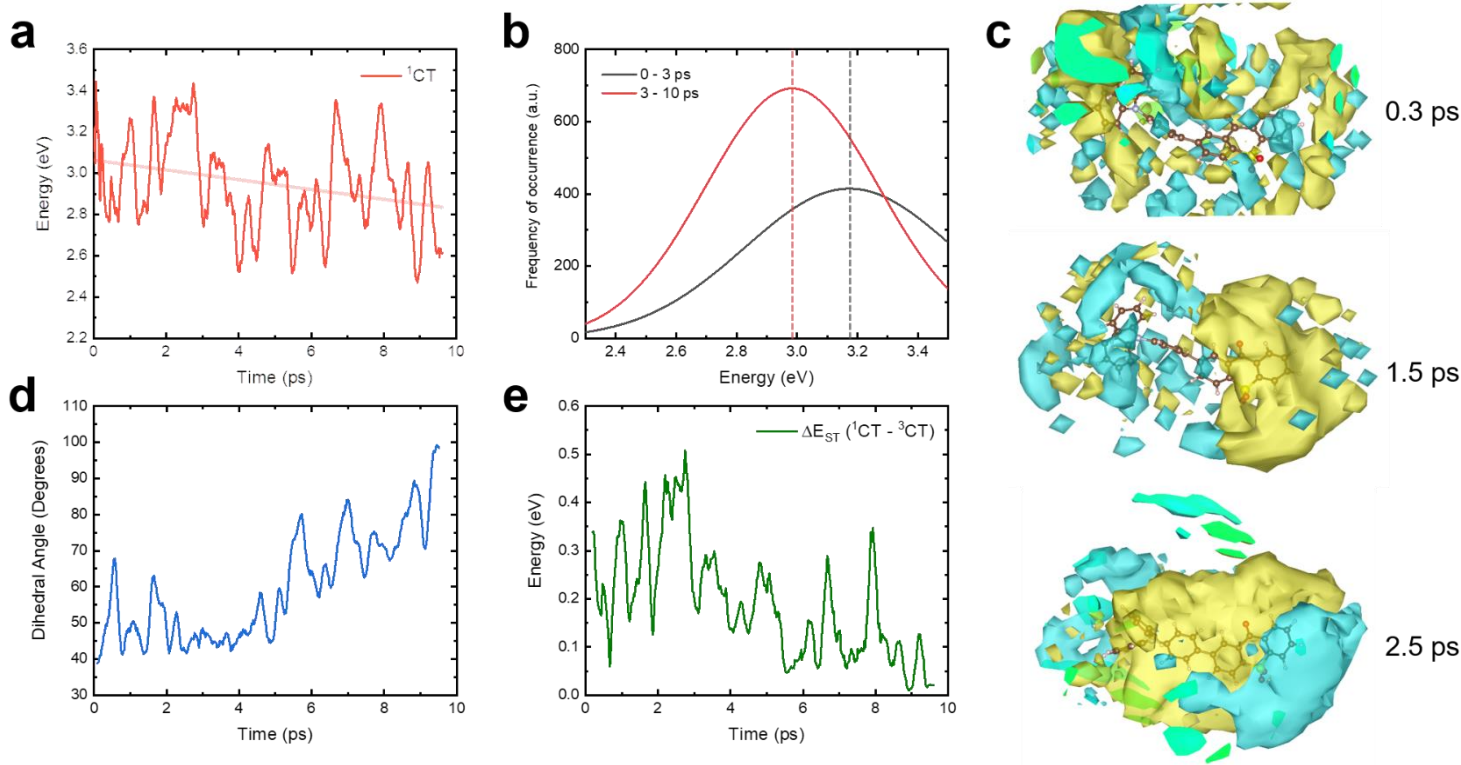

**Figure 4:** (**a**) The vertical excitation energy of the $^1$CT state along the simulation trajectory. The overlaid line is a linear best-fit guide to the eye. (**b**) The distribution of the vertical $^1$CT excitation energies during early (0-3 ps) and late (3-10 ps) sections of the simulation trajectory. The dotted lines overlaid mark the peak of the $^1$CT energy distribution, showing that the average energy of the $^1$CT state decreases due to the environment reorganisation. (**c**) A visualisation of the electrostatic potential of the solvent experienced by TXO-TPA at the indicated timescales of the trajectory. By 2.5 ps, the toluene molecules surrounding TXO-TPA are strongly polarised in response to the optical formation of the highly polar $^1$CT state. (**d**) The evolution of the D:A dihedral angle over the simulation trajectory. The dihedral angle increases from ~50º to ~90º after 4-5 ps, indicating the formation of a TICT state. (**e**) The $\Delta E_{ST}$ between the $^1$CT and $^3$CT states along the simulation trajectory.
21

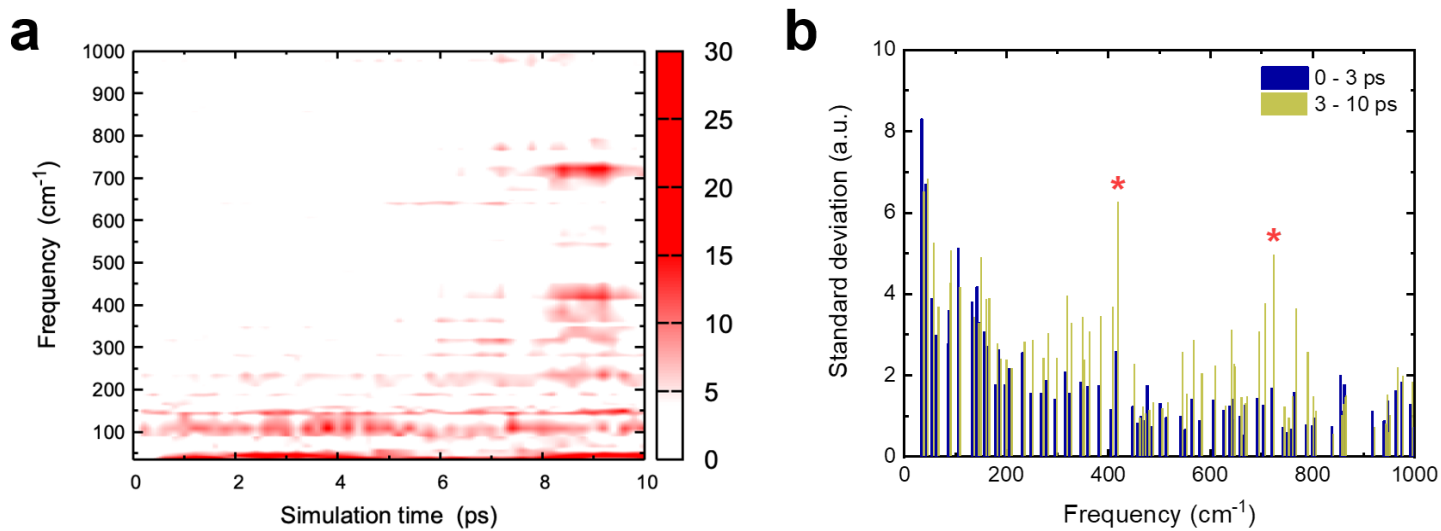

**Figure 5: (a)** The time evolution of the normal modes along the $^1$CT adiabatic dynamics. The colour scale indicates the mode intensity. **(b)** The standard deviations of the mode intensities computed during the early (0-3 ps) and late (3-10 ps) simulation timescales. Over the later timescales where the solvent has reorganised to drive the formation of a quasi-pure CT state, the modes at 417 and 712 cm$^{-1}$ become strongly active (as denoted by the red asterisks).

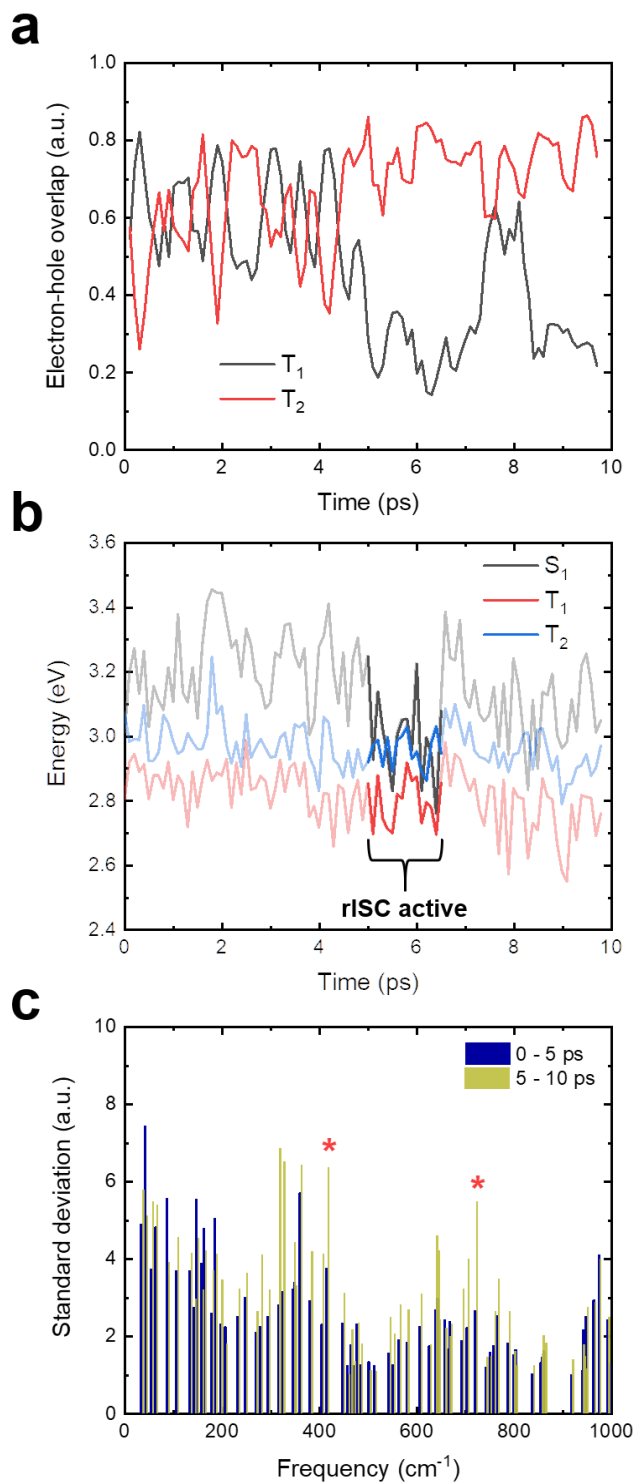

**Figure 6: (a)** The electron/hole overlap for the two lowest energy triplet states, indicating whether they possess primarily $^3$LE (large overlap) or $^3$CT (small overlap) character during the timescales of the simulation. **(b)** The energy of the $S_1$, $T_1$ and $T_2$ states over time timescales of the simulation. The bold section between 5-6.5 ps indicates the region where all three electronic excited states are confined within a narrow energy range, enabling rISC to take place. **(c)** The standard deviations of the mode intensities computed during the early (0-5 ps) and late (5-10 ps) simulation timescales. Over the later timescales where the solvent has reorganised to stabilise the $^3$CT state, clusters of modes around 400 and 700 cm$^{-1}$ become strongly active (as denoted by the red asterisks).

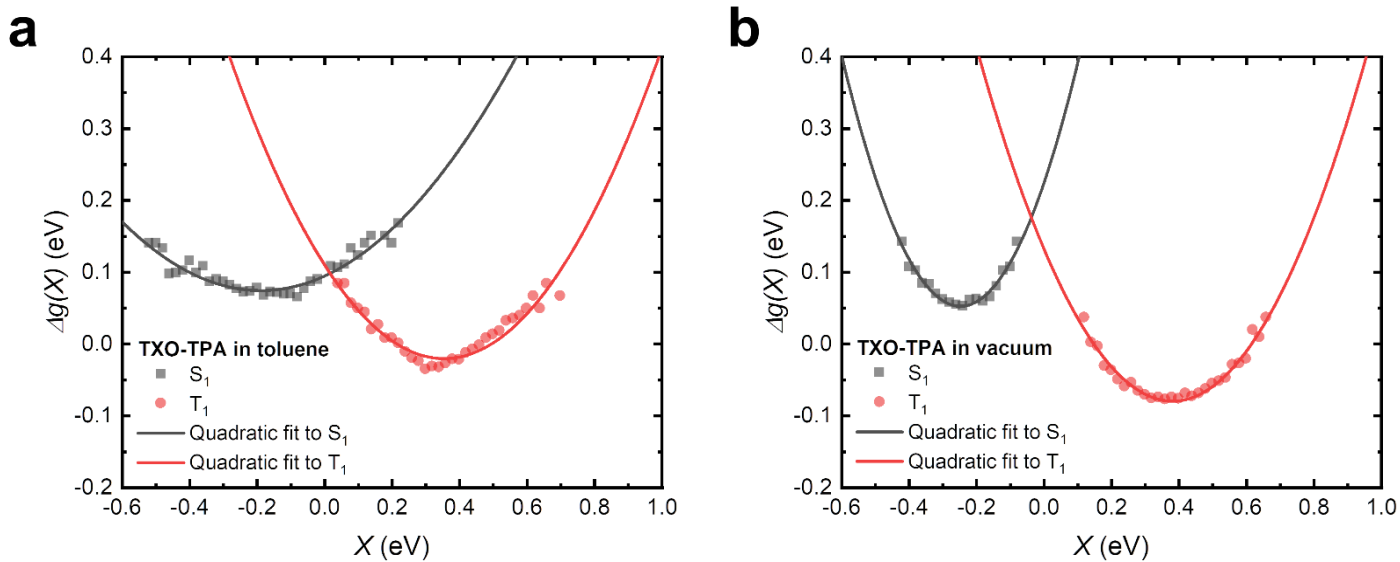

**Figure 7: (a)** Free energy functions for $S_1$ (back squares) and $T_1$ (red circles) states together with their quadratic fits for TXO-TPA in an explicit toluene solvent environment. **(b)** Free energy functions for $S_1$ (back squares) and $T_1$ (red circles) states together with their quadratic fits for TXO-TPA in vacuum.

# Methods

**Transient Absorption**

Transient absorption (TA) measurements were performed on a home-built TA setup, powered by a Ti:sapphire amplifier (Spectra Physics Solstice Ace) that generated 100 fs duration pulses centred at 800 nm, with a repetition rate of 1 kHz. On this setup, sample photoexcitation in the short-time TA experiments (100 fs-1.7 ns) was achieved by the second harmonic (400 nm) of the 800 nm output of the Ti:sapphire amplifier, generated using a β-Barium Borate (BBO) crystal and filtered with a 3 mm thick BG39 glass filter to remove the residual 800 nm fundamental pulse. The probe light was generated by home-built broadband visible (500-770 nm) and near-infrared (830-1025 nm) non-collinear optical parametric amplifiers (NOPAs), pumped using the 400 nm second harmonic of the Ti:sapphire output. For short-time measurements, a mechanical delay stage (Thorlabs DDS300-E/M) was used provide the pump-probe delay. In the long-time measurements, the delay between probe and pump pulses was varied using a Stanford DG645 delay generator. The transmitted probe pulses were collected with a silicon dual-line array detector (Hamamatsu S8381-1024Q) which was driven and read out by a custom-built board from Stresing Entwicklungsbüro.

**Transient Grating Photoluminescence**

A Ti:Sapphire amplifier system (Spectra Physics Solstice Ace) operating at 1 kHz generating 100 fs pulses was split into the pump and probe beam arms. The pump beam was generated by second harmonic generation (SHG) in a BBO crystal and focused onto the sample. The photoluminescence generated is collimated using a silver off-axis parabolic mirror and focused onto the gate medium. About 80 µJ/pulse of the 800 nm laser output is used for the gate beams, which is first raised 25 mm above the plane of the PL to produce a boxcar geometry and split into a pair of gate beams using a 50/50 beam splitter. The gate beams are focused onto the gate medium (fused silica), crossing at an angle of ∼5° and overlapping with the focused PL. The two gate beams interfere and create a transient grating in the gate medium due to a modulation of the refractive index via the optical Kerr effect. Temporal overlap between the two gate beams is achieved via a manual delay stage. The PL is then deflected on the transient grating causing a spatial separation of the gated signal from the PL background. Two lenses collimate and focus the gated signal onto the spectrometer entrance (Princeton Instruments SP 2150) after long- and short-pass filters remove scattered pump and gate light, respectively. Gated PL spectra are measured using an intensified CCD camera (Princeton Instruments, PIMAX4). The (∼10 ns) electronic shutter of the intensified CCD camera was used to further suppress long-lived PL background. PL spectra at each gate time delay are acquired from ∼10000 laser shots. The time delay between pump and gate beams is controlled via a motorized optical delay line on the excitation beam path and a LabVIEW data acquisition program.

**ns-µs Transient Photoluminescence**

Transient PL spectra were recorded using an electrically-gated intensified charge-coupled device (ICCD) camera (Andor iStar DH740 CCI-010) connected to a calibrated grating spectrometer (Andor SR303i). Pulsed 400 nm photoexcitation was provided at a repetition rate of 1 kHz. A 425 nm long-pass filter (Edmund Optics) was used to prevent scattered laser signal from entering the camera. Temporal evolution of the PL emission was obtained by stepping the ICCD gate delay with respect to the excitation pulse. The minimum gate width of the ICCD was 5 ns. Recorded data was subsequently corrected to account for filter transmission and camera sensitivity.

**Impulsive Vibrational Spectroscopy**

The IVS experiments were performed using a Yb:KGW laser system (Pharos, Light Conversion) to provide 15.2 W at 1030 nm with a 38 kHz repetition rate. The probe beam was generated by focusing a portion of the fundamental in a 4 mm YAG crystal to generate a white light continuum (WLC). The push beam was generated by a non-collinear optical parametric amplifier (NOPA) seeded by the WLC from a 3 mm YAG crystal mixed with a second harmonic pump (HIRO, Light Conversion) in a barium borate crystal (23.5º cut, type I phase matching, ~2º external angle). The NOPA output was compressed down to <10 fs pulses using a pair of chirped mirrors and pair of $CaF_2$ wedges, as determined by second-harmonic generation frequency-resolved optical gating (SHG-FROG) During the measurement, the push pulse intensity was kept low enough to avoid detectable two-photon absorption. The pump was generated using a narrow band optical parametric oscillator system (ORPHEUS- LYRA, Light conversion) with a 1030 nm seed. The FWHM of the pump was ~200 fs. The probe white light was delayed using a computer-controlled piezoelectric translation stage (Physik Instrumente). The pump was delayed using a computer controlled Thorlabs translation stage. The pulse train of probe pulses with/without the pump and with/without the push was generated using separate chopper wheels on the pump and push beams. After the sample, the probe pulse was split with a 950 nm dichroic mirror (Thorlabs). The visible part was then imaged with a Silicon photodiode array camera (Stresing Entwicklunsbüro; visible monochromator 550 nm blazed grating). The near-infrared portion of the probe was focused into to an IR spectrograph (1200 nm blazed grating) and imaged on an InGaAs photodiode array camera (Sensors Unlimited). Offsets for the differing spectral response of the detectors was accounted for in the post-processing of data.

Fast Fourier transforming the femtosecond-transient absorption data allows determination of the mode frequency modulating the electronic transition. Briefly, each row of the TA data, *y*, contains kinetic information at wavelength *i*, and is fitted with a multiple exponential decay functions.

$$y_i = \sum_n \frac{1}{2} \cdot A_n \cdot e^{\frac{s^2}{2\tau_n^2}} \cdot e^{\frac{-(t-t_0)}{\tau_n}} \cdot \left(1 + erf\left(\frac{t-t_0-\frac{s^2}{\tau_n}}{\sqrt{2}s}\right)\right) \quad (1)$$

The *n* lifetimes, τ, are treated as a common fitting parameter for the whole data set, each with a different weighting $A_n$ at a particular wavelength. A chirp and background correction were performed before the global fit.

$$y_i = \sum_n \frac{1}{2} \cdot A_n \cdot e^{\frac{s^2}{2\tau_n^2}} \cdot e^{\frac{-(t-t_0)}{\tau_n}} \cdot \left(1 + erf\left(\frac{t-t_0-\frac{s^2}{\tau_n}}{\sqrt{2}s}\right)\right) \quad (2)$$

The error function considers the Gaussian-shaped probe pulse arriving at t = $t_0$ and instrumental response time, *s*. The TA data can be modelled as a matrix multiplication in the form of

$$\mathbf{y} = \mathbf{A} \cdot \mathbf{E} = \begin{pmatrix} A_{11} & \cdots & A_{1n} \\ \vdots & \ddots & \vdots \\ A_{m1} & \cdots & A_{mn} \end{pmatrix} \cdot \begin{pmatrix} E_{11} & \cdots & E_{1t} \\ \vdots & \ddots & \vdots \\ E_{n1} & \cdots & E_{nt} \end{pmatrix} \quad (3)$$

where *m, n* and *t* are the number of wavelengths, lifetimes and timepoints in the data. **E** is a matrix containing all the exponential components in Equation 2 and **A** can be computed by a simple matrix division provided we have initial guesses for the various parameters. τ, $t_0$ and *s* were then fitted iteratively, and **A** updated after each iteration, until convergence was achieved.

A Blackman window was then applied to the residual map before Fast Fourier Transform was performed on the linear time region. The data was padded with zero arrays to improve FFT performance.

**Steady-state absorption**

Steady-state absorption spectra were measured using an HP 8453 spectrometer.

**Steady-state photoluminescence**

The steady-state PL was measured in an integrating sphere, where the samples were excited with a 405 nm continuous-wave laser. The emission signals were measured using a calibrated Andor iDus DU420A BVF Si detector.

**Photoluminescence quantum efficiency measurements**

The PLQE was determined using method previously described by De Mello *et al.*[50]. Samples were placed in an integrating sphere and photoexcited using a 405 nm continuous-wave laser. The laser and emission signals were measured and quantified using a calibrated Andor iDus DU420A BVF Si detector.

**Raman spectroscopy**

Raman spectroscopy measurements were performed on films fabricated on a quartz substrate. The Raman spectra were collected using a HORIBA T64000 Raman spectrometer attached to a confocal microscope with a 100× objective and a 633 nm laser for the excitation. Laser power was minimized to ensure that no degradation of the sample was induced. The spectra were averaged over several accumulations.

**Computational details**

The changes in dipole moment upon a vertical excitation to the lowest singlet state were calculated at the density functional theory (DFT) level using the PBEh-3c functional[51], implemented in ORCA software[52]. To study the time-evolution of the singlet and triplet states, our computational modelling uses a combined QM/MM approach. For these investigations, a single TXO-TPA molecule was surrounded by 249 individual toluene molecules and the reorganization dynamics following population of the $^1$CT and $^3$CT states were determined. This was achieved by employing an electrostatic embedding scheme, where interaction with the solvent includes the electronic polarization of the QM TXO-TPA molecule by re-orientation of the MM toluene molecules, together with Lennard-Jones potentials. Thus, this approach accounts for the (slow) inertial component of the solvent dielectric response, whilst the optical component that produces an instantaneous and rigid shift of the transition energies is ignored[53]. The TXO-TPA molecule was described at the DFT level, while the surrounding 249 toluene molecules were treated with the GROMOS 54a7 classical force field[54] in Gomacs code[55]. We explored the adiabatic excited-state dynamics in the $^1$CT for the first 10 ps using the forces from time-dependent (TD)-DFT. The same approach was further utilized to study the triplet dynamics, sampling the geometries from 10 ps of adiabatic dynamics in the lowest triplet state from open-shell DFT calculations, before applying spin-adapted TD-DFT calculations to access the properties of higher-lying triplet excitons. The electron and hole densities of the three lowest energy triplet states ($T_1$, $T_2$ and $T_3$), calculated in the gas-phase ground-state equilibrium geometry of TXO-TPA, are displayed in Fig. S25. We note that whilst $T_2$ and $T_3$ are $^3$LE states localized on the TXO moiety, $T_1$ has primarily $^3$CT character. Finally, to quantify the extension of the normal modes along the molecular backbone, we defined the participation ratio as $PR(D/A) = \sum \Delta_{D/A}^2 / \sum \Delta^2$, where $\Delta$ is a Cartesian displacement for each atom. Further computational details are provided in SI.

**Acknowledgements**

A.J.G. and R.H.F. acknowledge support from the Simons Foundation (grant no. 601946) and the EPSRC (EP/M01083X/1 and EP/M005143/1). A.P., Y.O. and D.B. were supported by the European Union's Horizon 2020 research and innovation programme under Marie Sklodowska Curie Grant agreement 748042 (MILORD project). Computational resources in Mons were provided by the Consortium des Équipements de Calcul Intensif (CÉCI), funded by the Fonds de la Recherche Scientifiques de Belgique (F.R.S.-FNRS) under Grant No. 2.5020.11, as well as the Tier-1 supercomputer of the Fedération Wallonie-Bruxelles, infrastructure funded by the Walloon Region under Grant Agreement No. 1117545. D.B. is a FNRS Research Director. This project has received funding from the European Research Council (ERC) under the European Union's Horizon 2020 research and innovation programme (grant agreements No.'s 670405 and 758826). We thank the Winton Programme for the Physics of Sustainability and the Engineering and Physical Sciences Research Council for funding. R.P. acknowledges financial support from an EPSRC Doctoral Prize Fellowship. A.J.S acknowledges the Royal Society Te Apārangi and the Cambridge Commonwealth European and International Trust for their financial support. Y.O. acknowledges funding by the Fonds de la Recherche Scientifique-FNRS under Grant no. F.4534.21 (MIS-IMAGINE). We thank C. Schnedermann for useful discussions.

**Author contributions**

A.J.G. and A.P. contributed equally. A.J.G, R.H.F. and D.B. conceived the work. A.J.G. performed the TA measurements. A.J.G., R.P and A.J.S. carried out the IVS measurements. S.F. performed the TGPL measurements. E.W.E., T.H.T. and B.H.D. carried out the trPL measurements. T.H.T. performed the Raman spectroscopy. A.J.G., E.W.E and L.-S.C. fabricated the samples used in the work. A.P. carried out the quantum chemical calculations.

A.M.A., G.D.S, Y.O. and D.B. discussed the calculation results. A.R., R.H.F. and D.B. supervised their group members involved in the project. A.J.G., A.P., G.D.S., R.H.F. and D.B. wrote the manuscript with input from all authors.

## Competing financial interests

The authors declare no competing interests.

## Additional information

Supplementary information accompanies this paper at [to be completed in proofs].

Correspondence and requests for materials should be addressed to D.B. (david.beljonne@umons.ac.be) and A.J.G. (ajg216@cam.ac.uk).

## Data availability

The data that supports the plots within this paper is available at the University of Cambridge Repository at [to be completed in proofs].

34  M. Kumar and L. Pereira, *Nanomaterials*, 2020, **10**, 101.

35  V. Karunakaran and S. Das, *J. Phys. Chem. B*, 2016, **120**, 7016–7023.

36  J. Choi, D.-S. Ahn, K. Y. Oang, D. W. Cho and H. Ihee, *J. Phys. Chem. C*, 2017, **121**, 24317–24323.

37  A. Petrozza, F. Laquai, I. A. Howard, J.-S. Kim and R. H. Friend, *Phys. Rev. B*, 2010, **81**, 205421.

38  T. Hosokai, H. Matsuzaki, H. Nakanotani, K. Tokumaru, T. Tsutsui, A. Furube, K. Nasu, H. Nomura, M. Yahiro and C. Adachi, *Sci. Adv.*, 2017, **3**, e1603282.

39  O. V. Mikhnenko, P. W. M. Blom and T.-Q. Nguyen, *Energy Environ. Sci.*, 2015, **8**, 1867–1888.

40  M. Liebel and P. Kukura, *J. Phys. Chem. Lett.*, 2013, **4**, 1358–1364.

41  D. Di, A. S. Romanov, L. Yang, J. M. Richter, J. P. H. Rivett, S. Jones, T. H. Thomas, M. Abdi Jalebi, R. H. Friend, M. Linnolahti, M. Bochmann and D. Credgington, *Science (80-. ).*, 2017, **356**, 159–163.

42  S. Sasaki, G. P. C. Drummen and G. I. Konishi, *J. Mater. Chem. C*, 2016, 4, 2731–2743.

43  P. L. Santos, J. S. Ward, P. Data, A. S. Batsanov, M. R. Bryce, F. B. Dias and A. P. Monkman, *J. Mater. Chem. C*, 2016, **4**, 3815–3824.

44  S. Rafiq, B. Fu, B. Kudisch and G. D. Scholes, *Nat. Chem.*, 2021, **13**, 70–76.

45  N. Aizawa, Y. Harabuchi, S. Maeda and Y.-J. Pu, *Nat. Commun.*, 2020, **11**, 3909.

46  Y. Niu, Q. Peng and Z. Shuai, *Sci. China Ser. B Chem.*, 2008, **51**, 1153–1158.

47  A. Warshel and W. W. Parson, *Annu. Rev. Phys. Chem.*, 1991, **42**, 279–309.

# Supplementary information for

# Dielectric control of reverse intersystem crossing in thermally-activated delayed fluorescence emitters

*Alexander J. Gillett[1*#], Anton Pershin[2,3#], Raj Pandya[1], Sascha Feldmann[1], Alexander J. Sneyd[1], Antonios M. Alvertis[1], Emrys W. Evans[1,4], Tudor H. Thomas[1], Lin-Song Cui[1], Bluebell H. Drummond[1], Gregory D. Scholes[5], Yoann Olivier[2,6], Akshay Rao[1], Richard H. Friend[1] and David Beljonne[2*].*

[1]Cavendish Laboratory, University of Cambridge, JJ Thomson Avenue, Cambridge, CB3 0HE, UK.

[2]Laboratory for Chemistry of Novel Materials, Université de Mons, Place du Parc 20, 7000 Mons, Belgium.

[3]Wigner Research Centre for Physics, PO Box 49, H-1525, Budapest, Hungary.

[4]Department of Chemistry, Swansea University, Singleton Park, Swansea, SA2 8PP, UK.

[5]Department of Chemistry, Princeton University, Princeton, New Jersey 08544, United States.

[6]Unité de Chimie Physique Théorique et Structurale & Laboratoire de Physique du Solide, Namur Institute of Structured Matter, Université de Namur, Rue de Bruxelles, 61, 5000 Namur, Belgium.

[*]Corresponding authors: David Beljonne: E-mail: david.beljonne@umons.ac.be; Alexander J. Gillett: E-mail: ajg216@cam.ac.uk.

[#]These authors contributed equally.

# Table of Contents

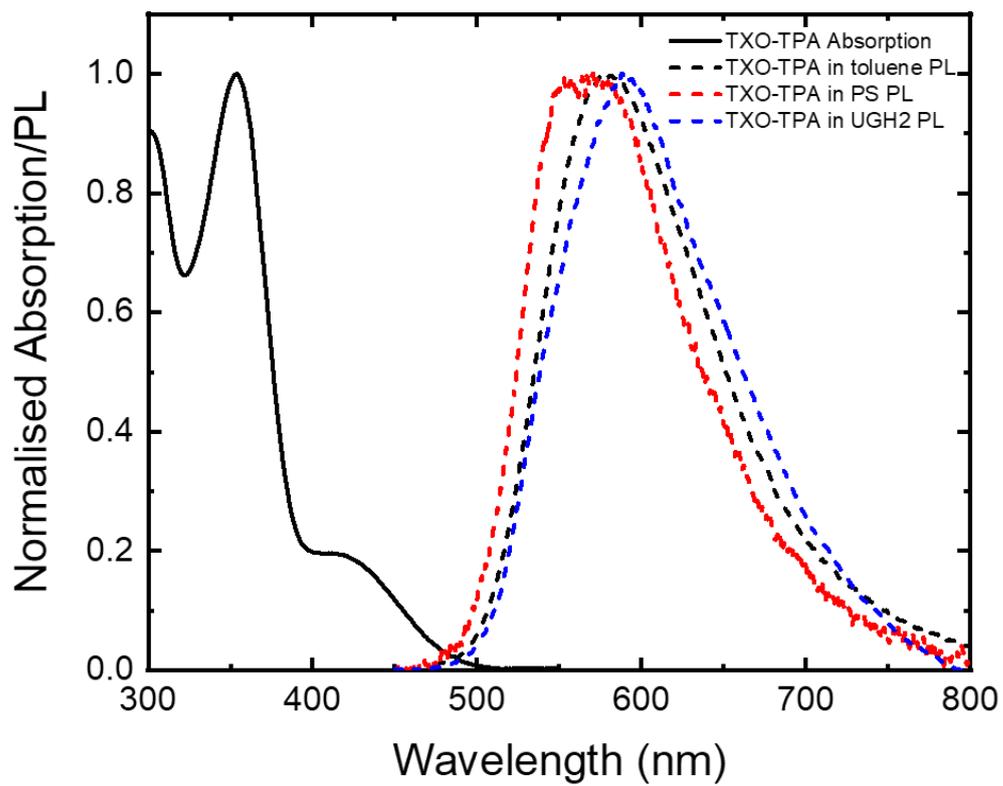

**Figure S1:** The absorption spectrum of TXO-TPA in a toluene solution and the emission spectra of TXO-TPA in a toluene solution, doped at 5 wt% into polystyrene and doped at 10 wt% into UGH2.

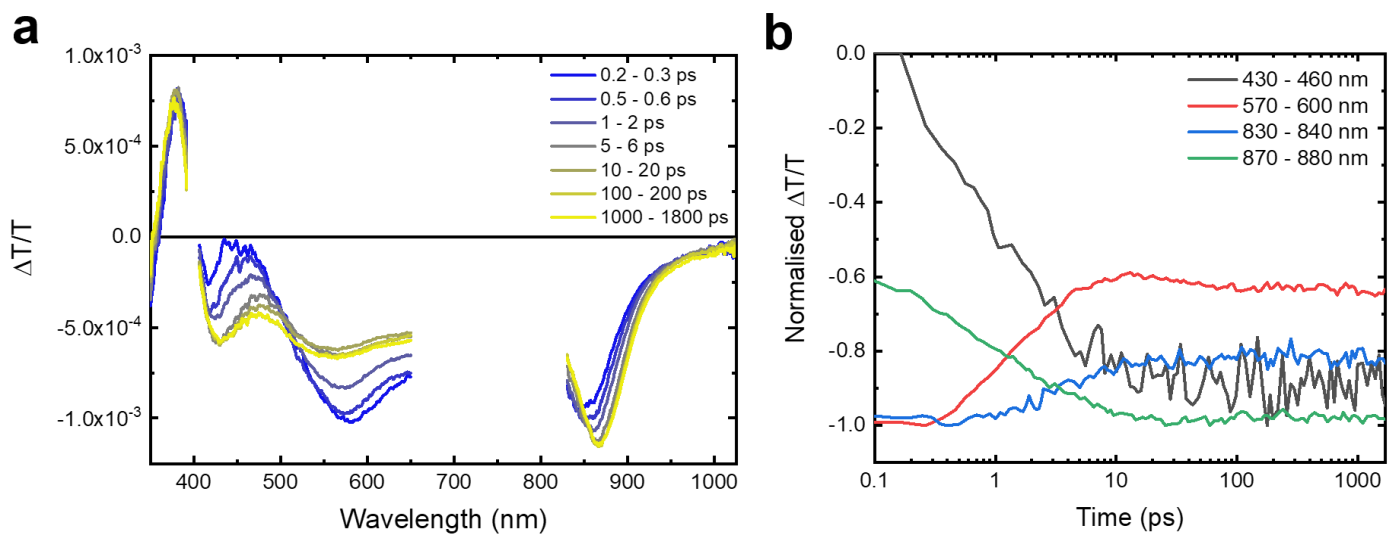

**Figure S2:** **(a)** The TA spectra of 2CzPN in a toluene solution, excited at 400 nm with a fluence of 28.3 μJ/cm$^2$. **(b)** The TA kinetics of 2CzPN in toluene.

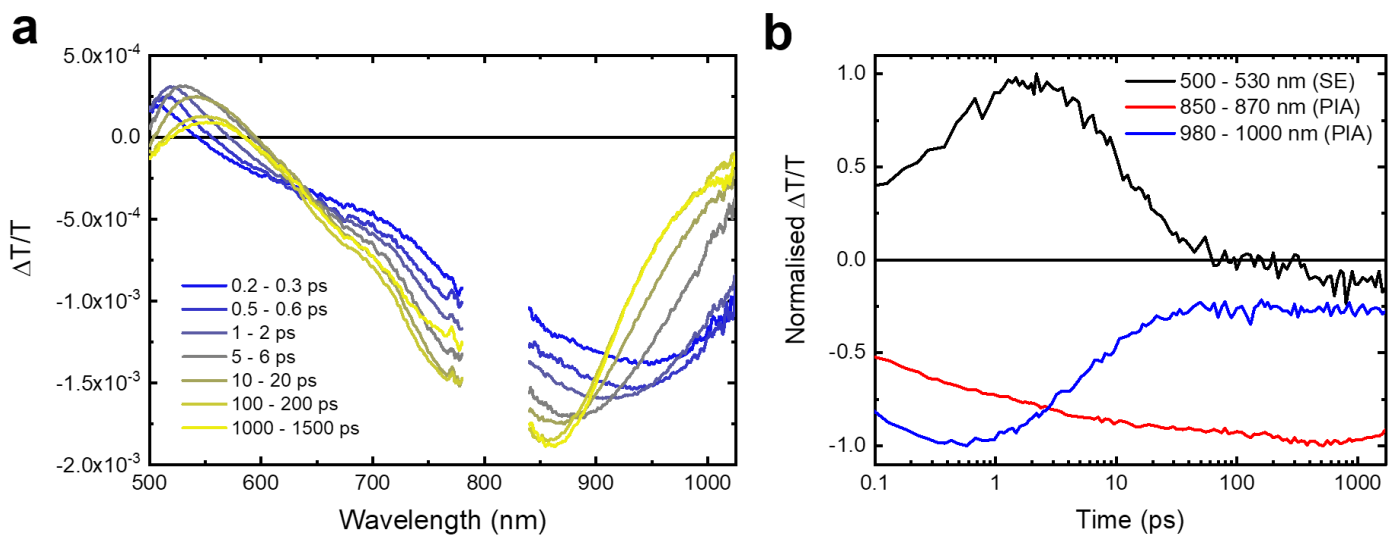

**Figure S3:** **(a)** The TA spectra of DACT-II in a toluene solution, excited at 400 nm with a fluence of 4.0 μJ/cm$^2$. **(b)** The TA kinetics of DACT-II in toluene.

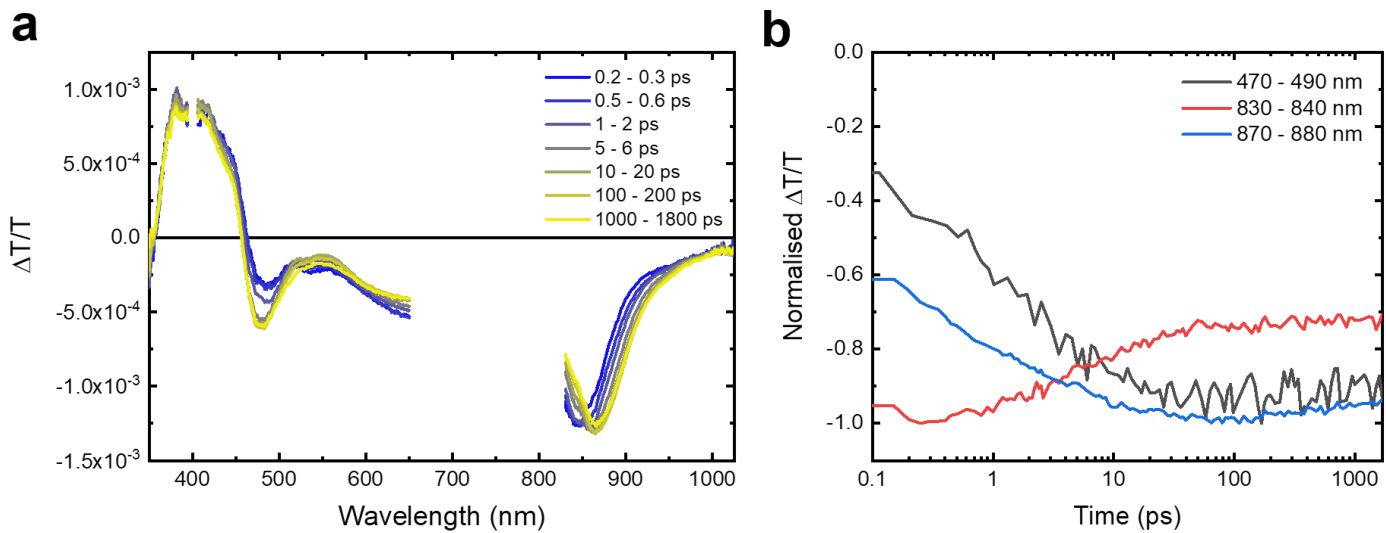

**Figure S4: (a)** The TA spectra of 4CzIPN in a toluene solution, excited at 400 nm with a fluence of 28.3 μJ/cm$^2$. **(b)** The TA kinetics of 4CzIPN in toluene.

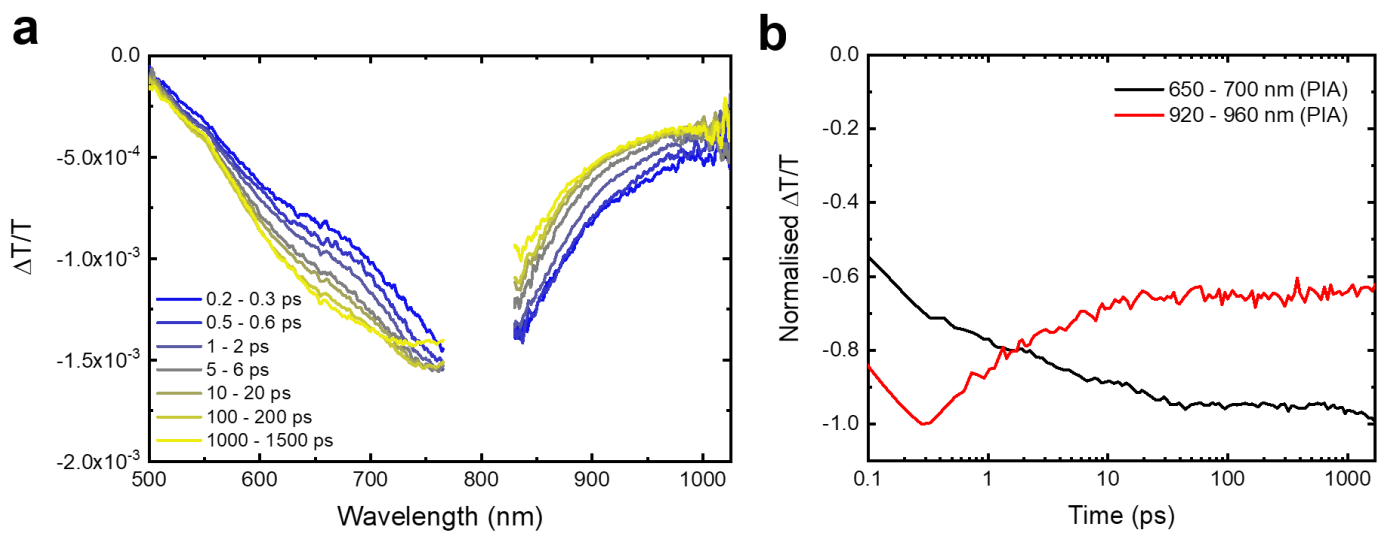

**Figure S5: (a)** The TA spectra of 5Cz-TRZ in a toluene solution, excited at 400 nm with a fluence of 4.0 μJ/cm$^2$. **(b)** The TA kinetics of 5Cz-TRZ in toluene.

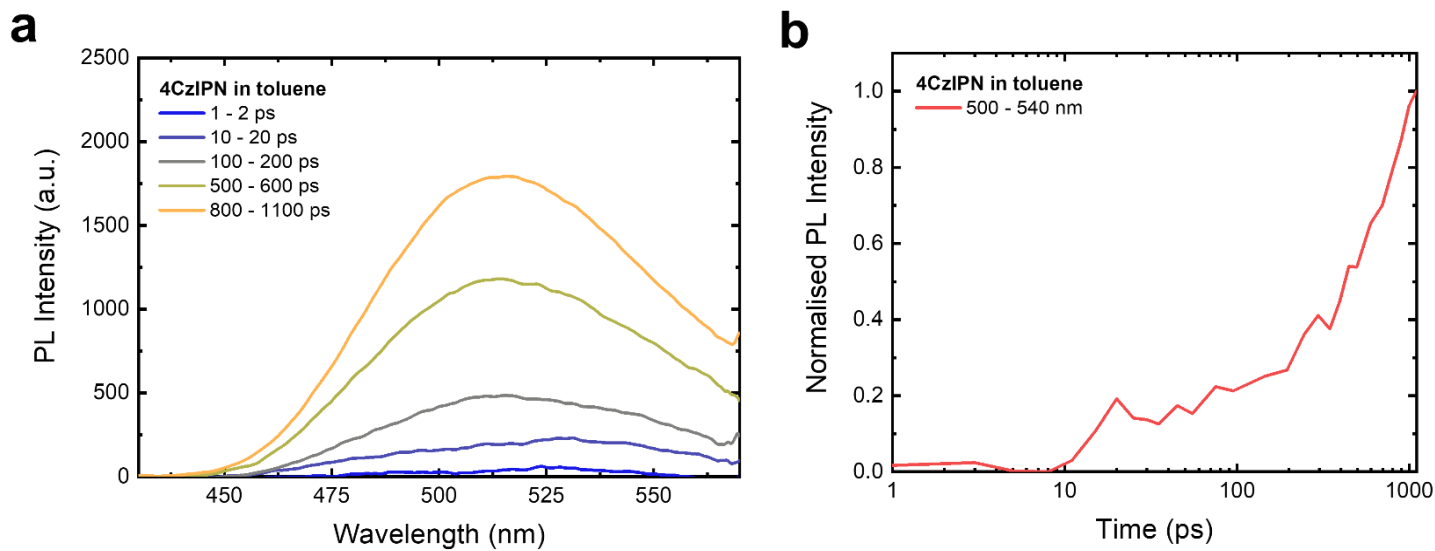

**Figure S6: (a)** The TGPL spectra of 4CzIPN in a toluene solution, excited at 400 nm with a fluence of 50.9 μJ/cm$^2$. **(b)** The kinetics of the TGPL, taken around the peak of the 4CzIPN PL (500-540 nm).

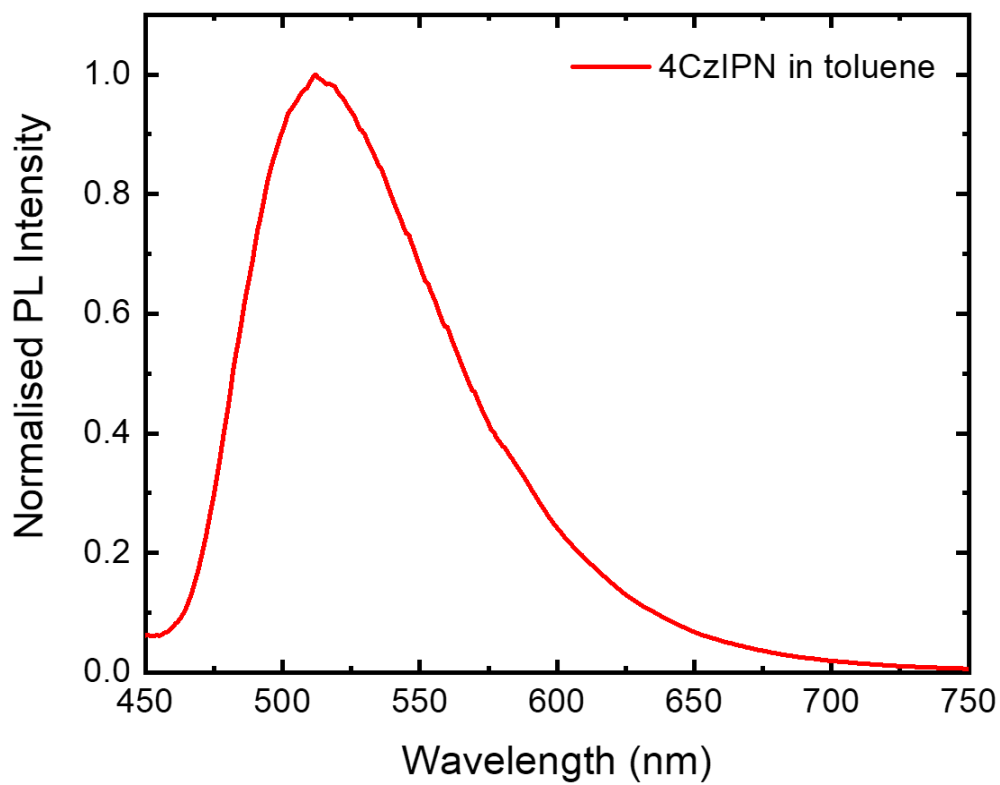

**Figure S7:** The steady-state PL spectra of 4CzIPN in a toluene solution, excited at 405 nm.

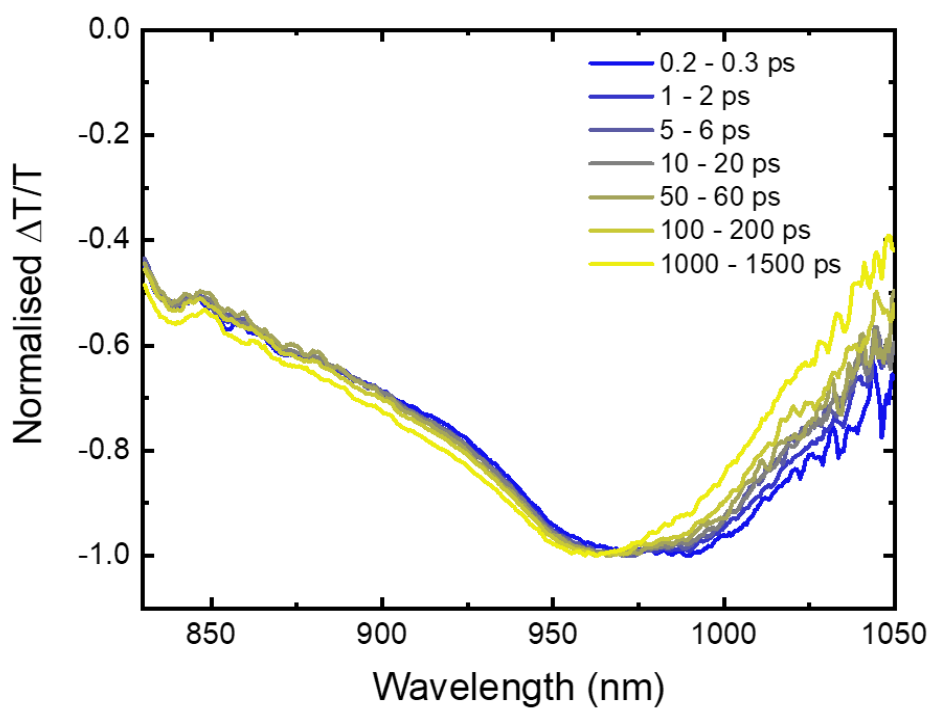

**Figure S8:** The normalized TA spectra of TXO-TPA in polystyrene (5 wt%), excited at 400 nm with a fluence of 7.8 μJ/cm$^2$. The film was encapsulated in an oxygen-free environment. The gradual blue-shift of the $^1$CT PIA over ps timescales is visible, though not as pronounced as in the toluene solution, reflecting the more constrained nature of the solid-state environment.

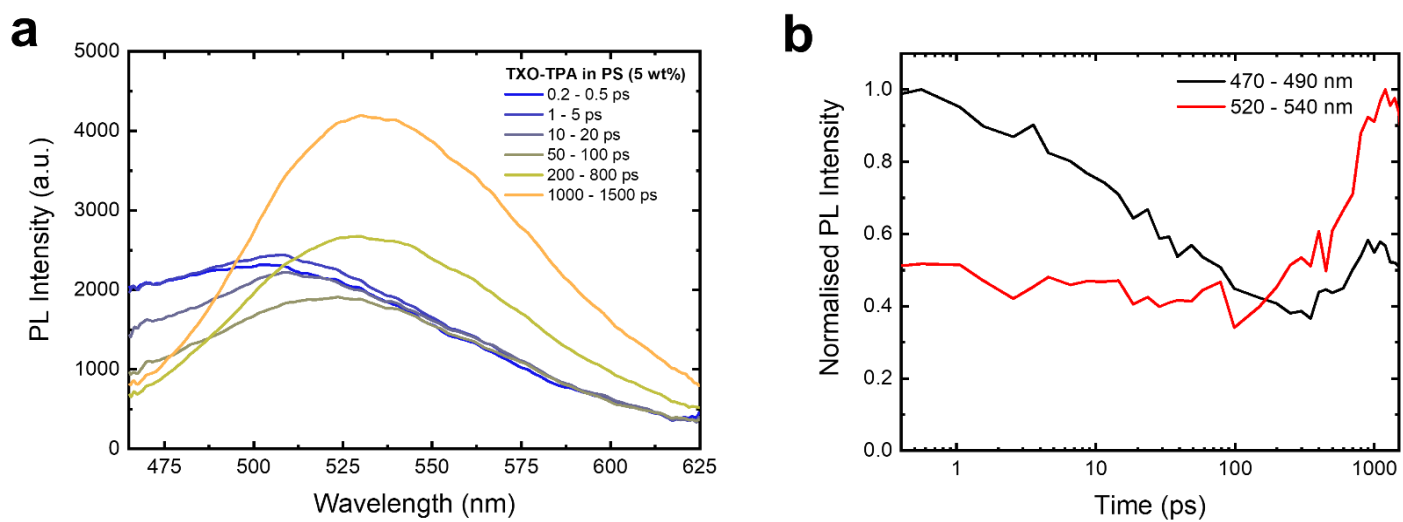

**Figure S9: (a)** The TGPL spectra of TXO-TPA in polystyrene (5 wt%), excited at 400 nm with a fluence of 50.9 μJ/cm$^2$. **(b)** The kinetics of the TGPL, taken at the high energy (470-490 nm) and low energy (520-540 nm) edges of the TXO-TPA PL. The PL maxima clearly red-shifts from 510 nm to 540 nm but does not reach the steady-state peak position of 570 nm.

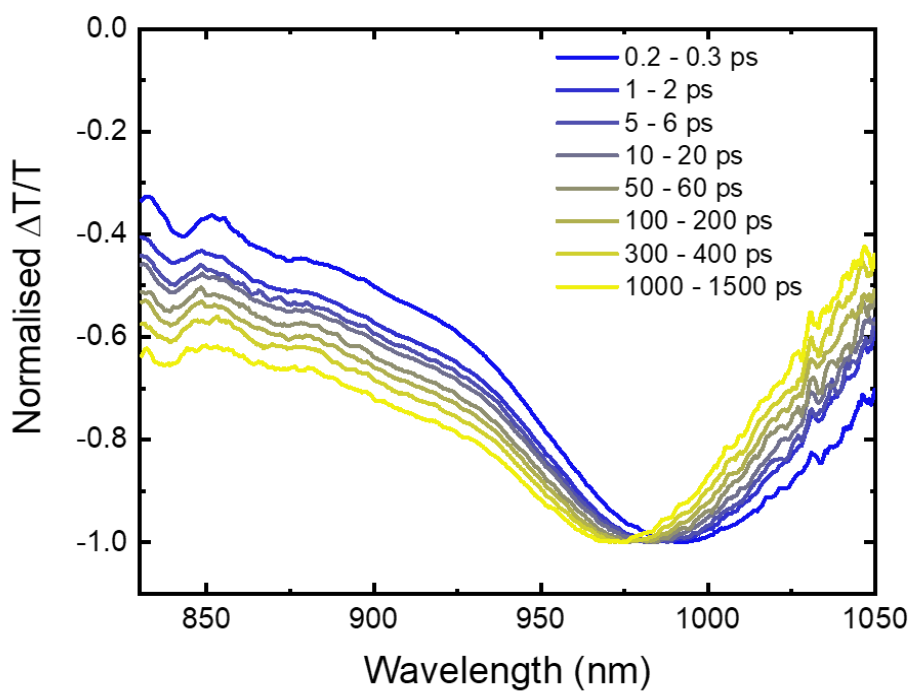

**Figure S10:** The normalized TA spectra of TXO-TPA in UGH2 (10 wt%), excited at 400 nm with a fluence of 44.6 μJ/cm$^2$. The film was encapsulated in an oxygen-free environment. The gradual blue-shift of the $^1$CT PIA over ps timescales is visible, though not as pronounced as in the toluene solution, reflecting the more constrained nature of the solid-state environment.

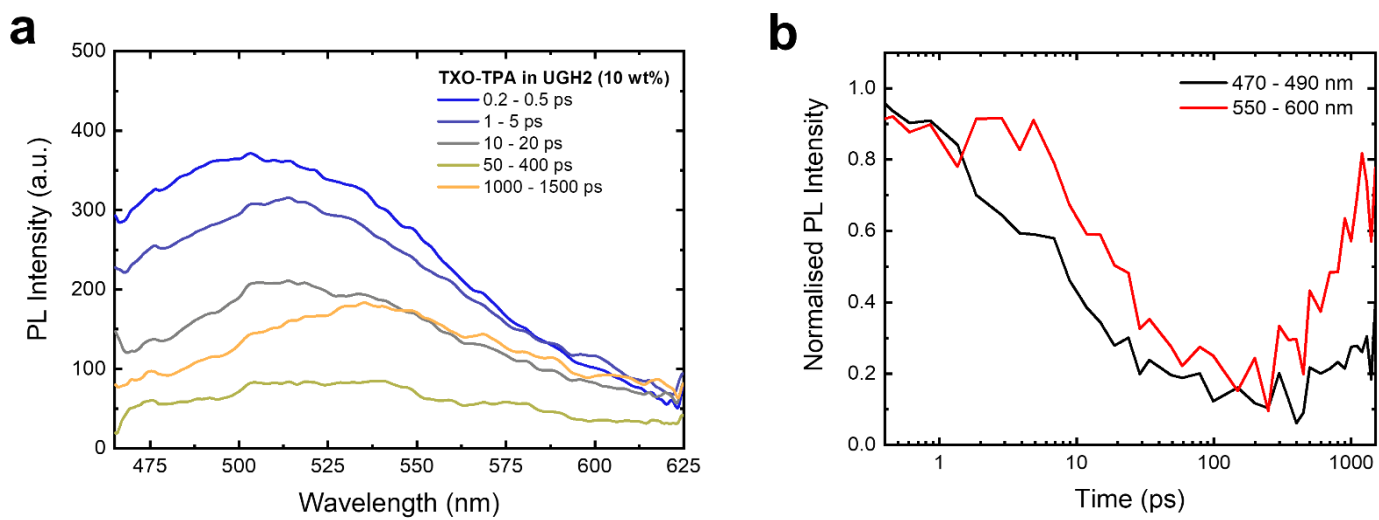

**Figure S11: (a)** The TGPL spectra of TXO-TPA in UGH2 (10 wt%), excited at 400 nm with a fluence of 50.9 μJ/cm$^2$. **(b)** The kinetics of the TGPL, taken at the high energy (470-490 nm) and low energy (550-600 nm) edges of the TXO-TPA PL. The PL maxima clearly red-shifts from 500 nm to 540 nm, but does not reach the steady-state peak position of 590 nm.

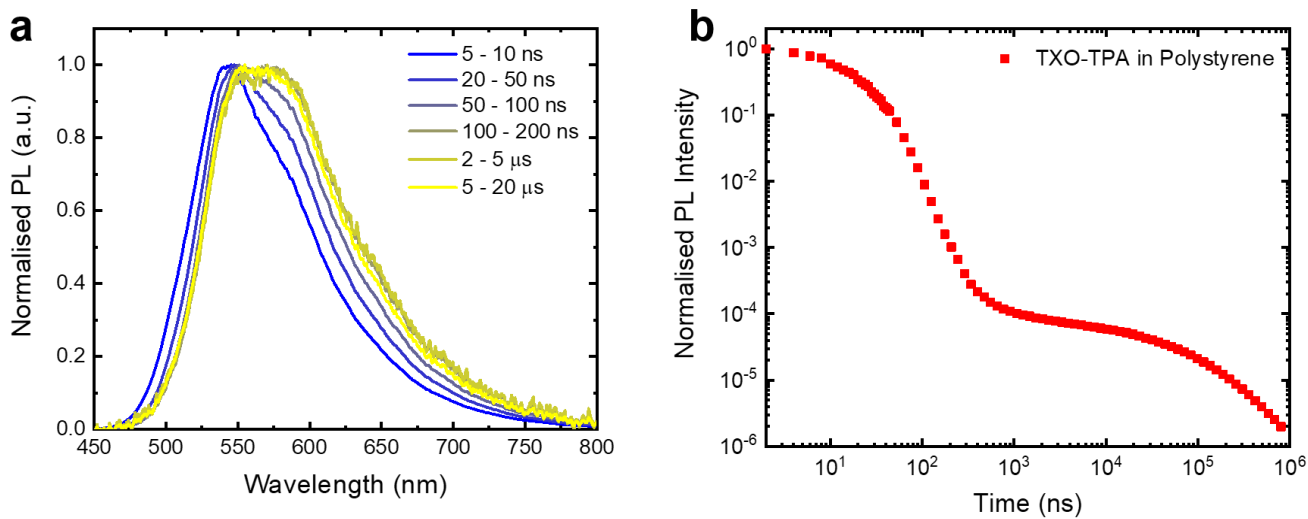

**Figure S12:** (a) The ns-μs trPL spectra of TXO-TPA in polystyrene (5 wt%), excited at 400 nm with a fluence of 10.0 μJ/cm$^2$. The PL peak shifts from 540 nm to the steady state maxima of 570 nm over timescales up to 100 ns. (b) The trPL kinetics of TXO-TPA, taken from the peak of the PL (560-600 nm).

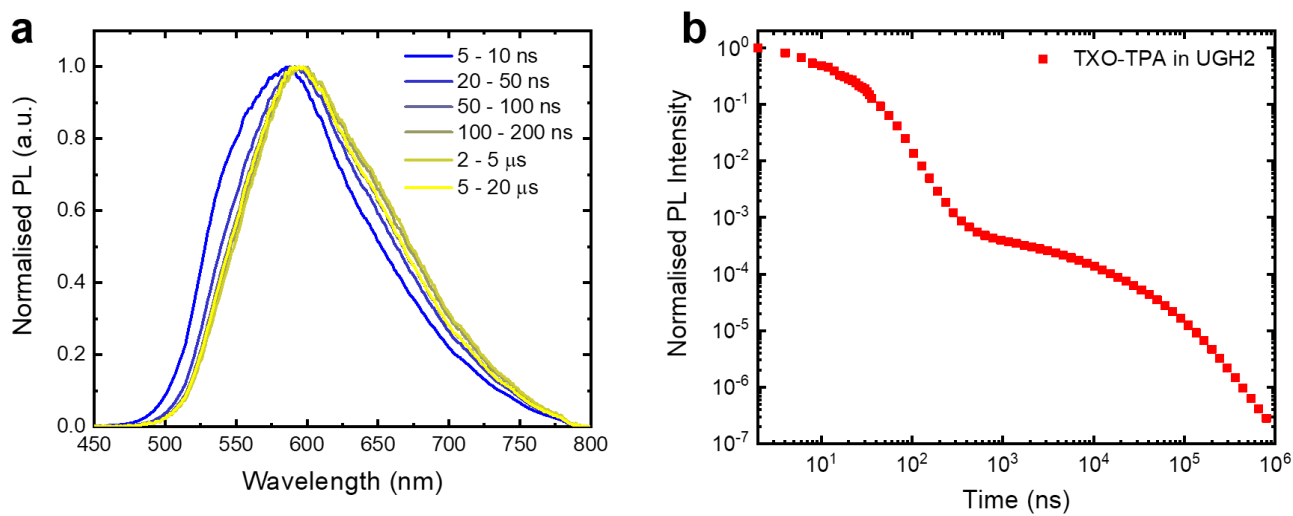

**Figure S13:** (a) The ns-μs trPL spectra of TXO-TPA in UGH2 (10 wt%), excited at 400 nm with a fluence of 10.0 μJ/cm$^2$. The PL peak shifts from 570 nm to the steady state maxima of 590 nm over timescales up to 20 ns. (b) The trPL kinetics of TXO-TPA, taken from the peak of the PL (580-610 nm).

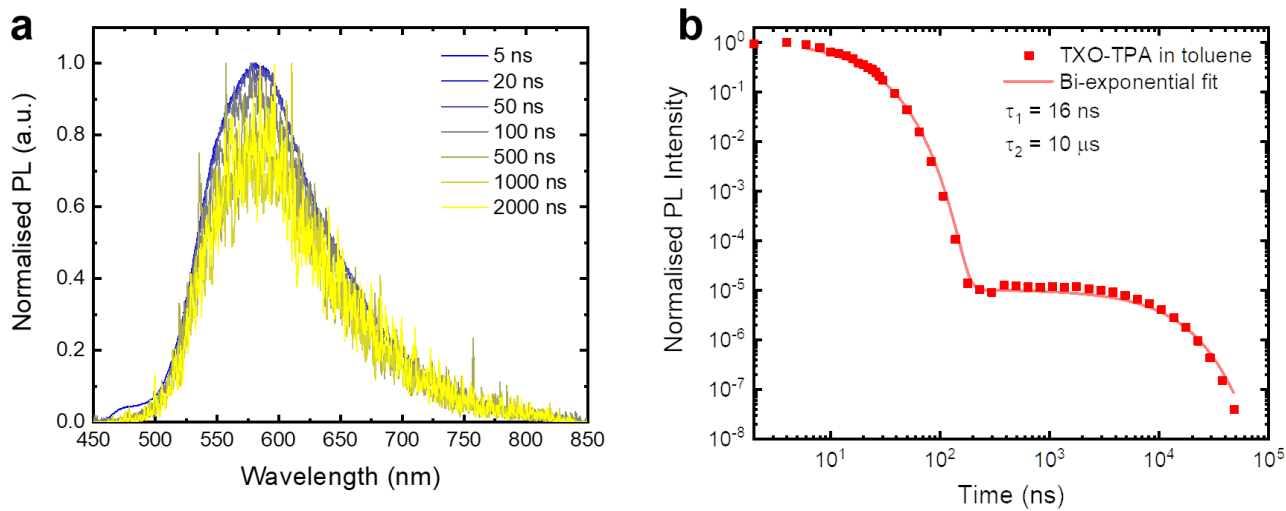

**Figure S14:** (a) The ns-μs trPL spectra of TXO-TPA in an oxygen-free toluene solution, excited at 400 nm with a fluence of 10.0 μJ/cm$^2$. No spectral shifts in the position of the PL maxima is observed over these timescales. (b) The trPL kinetics of TXO-TPA, taken from the peak of the PL (560-600 nm). The kinetic is fitted with a bi-exponential function, yield prompt and delayed time constants of 16 ns and 10 μs, respectively.

Using the experimental prompt and delayed lifetimes of 16 ns and 10 μs, respectively, as well as the photoluminescence quantum efficiency of TXO-TPA in toluene with (8.9%) and without (27.4%) oxygen exposure, we can determine the key photophysical rates from the following equations:

$$k_r^S = \phi_F k_p$$

$$k_{ISC} = (1 - \phi_F) k_p$$

$$k_{rISC} = \frac{k_p k_d}{k_{ISC}} \frac{\phi_{TADF}}{\phi_F}$$

$$k_{nr}^T = k_d - \phi_F k_{rISC}$$

From this, we obtain $k_r^S$ = 5.56x10$^6$ s$^{-1}$, $k_{ISC}$ = 5.69x10$^7$ s$^{-1}$, $k_{rISC}$ = 2.28x10$^5$ s$^{-1}$, and $k_{nr}^T$ = 7.97x10$^4$ s$^{-1}$ for TXO-TPA in toluene.

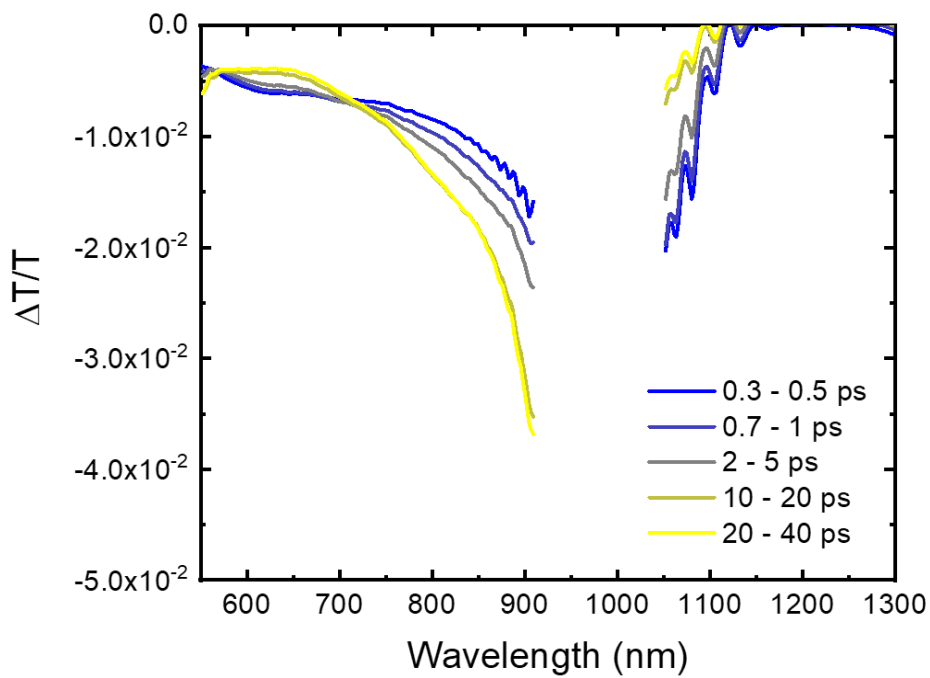

**Figure S15:** The TA spectra of TXO-TPA in a toluene solution, excited at 343 nm with a fluence of 0.75 mJ/cm$^2$. The gap in the 930-1050 nm probe region is due to the need to remove the intense fundamental of the laser at 1030 nm from the white light continuum used as the probe. The spectral evolutions that occur over the first ~20 ps closely resemble that of the TXO-TPA solution excited at 400 nm, confirming that the change in pump wavelength has negligible impact on the excited state dynamics of TXO-TPA.

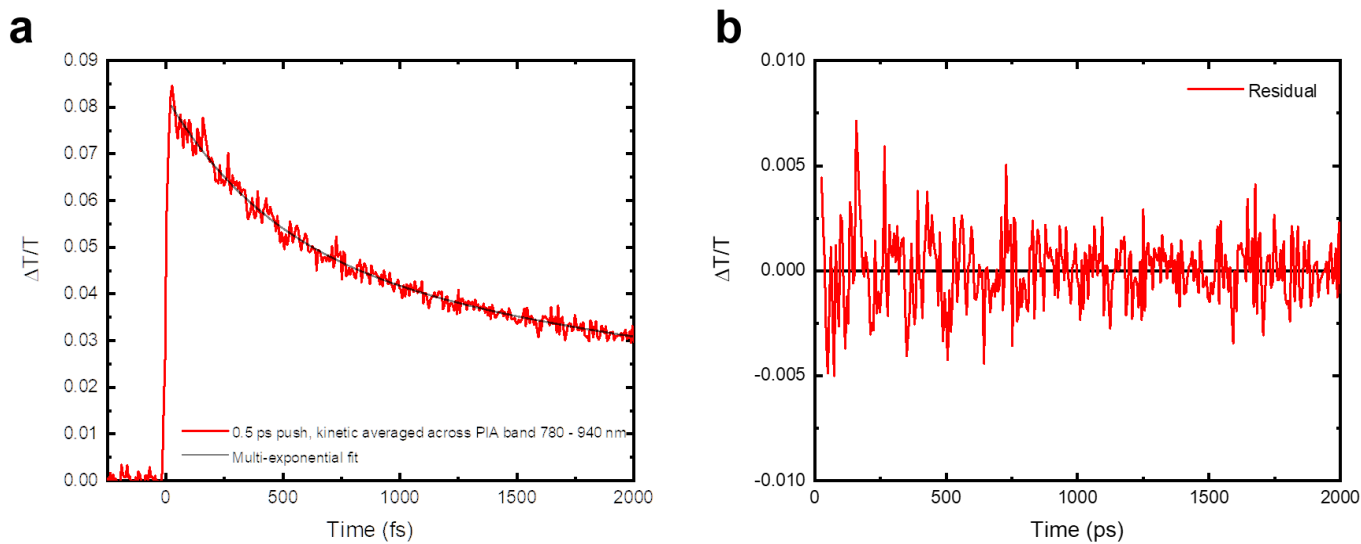

**Figure S16:** (a) The pump-push-probe kinetic taken from the $^1$CT PIA (780-940 nm) of TXO-TPA in a toluene solution, excited with a 270 fs 343 nm pulse and pushed 500 fs after the pump pulse with a 9.8 fs pulse centered at 790 nm, chosen to be resonant with the TXO-TPA $^1$CT PIA. The kinetic was fitted with a multi-exponential function to isolate the oscillations induced in the excited state by the fast push pulse. (b) The residual oscillations in the $^1$CT excited state extracted from the pump-push-probe kinetics. The vibrational modes associated with these oscillations were then extracted using a FFT.

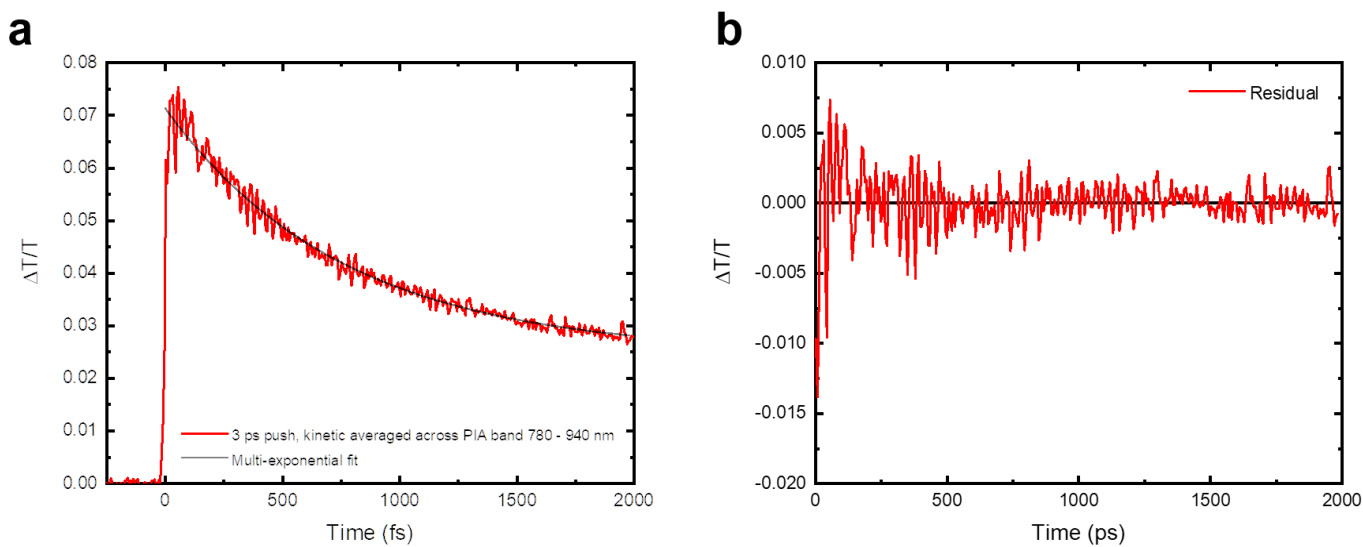

**Figure S17:** (a) The pump-push-probe kinetic taken from the $^1$CT PIA (780-940 nm) of TXO-TPA in a toluene solution, excited with a 270 fs 343 nm pulse and pushed 3 ps after the pump pulse with a 9.8 fs pulse centered at 790 nm, chosen to be resonant with the TXO-TPA $^1$CT PIA. The kinetic was fitted with a multi-exponential function to isolate the oscillations induced in the excited state by the fast push pulse. (b) The residual oscillations in the $^1$CT excited state extracted from the pump-push-probe kinetics. The vibrational modes associated with these oscillations were then extracted using a FFT.

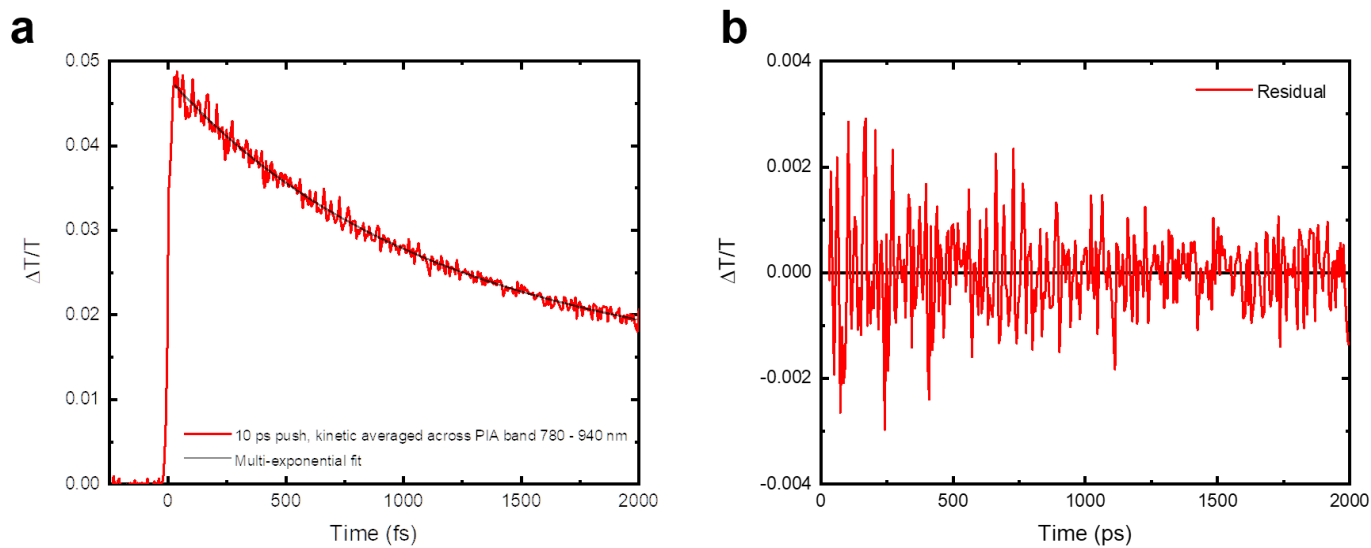

**Figure S18:** (a) The pump-push-probe kinetic taken from the $^1$CT PIA (780-940 nm) of TXO-TPA in a toluene solution, excited with a 270 fs 343 nm pulse and pushed 10 ps after the pump pulse with a 9.8 fs pulse centered at 790 nm, chosen to be resonant with the TXO-TPA $^1$CT PIA. The kinetic was fitted with a multi-exponential function to isolate the oscillations induced in the excited state by the fast push pulse. (b) The residual oscillations in the $^1$CT excited state extracted from the pump-push-probe kinetics. The vibrational modes associated with these oscillations were then extracted using a FFT.

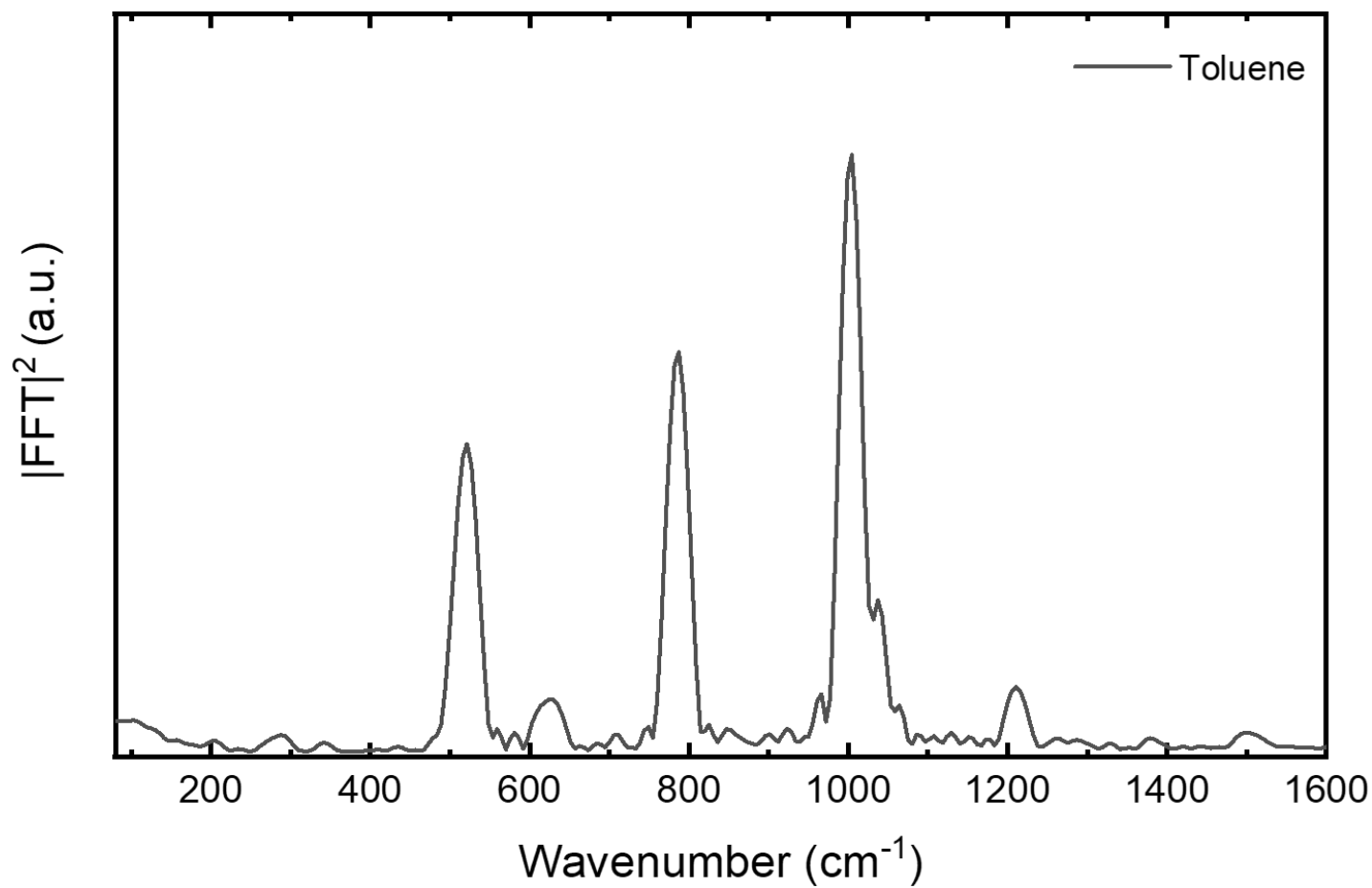

**Figure S19:** The IVS spectrum of toluene, measured under the same experimental conditions as the TADF emitters.

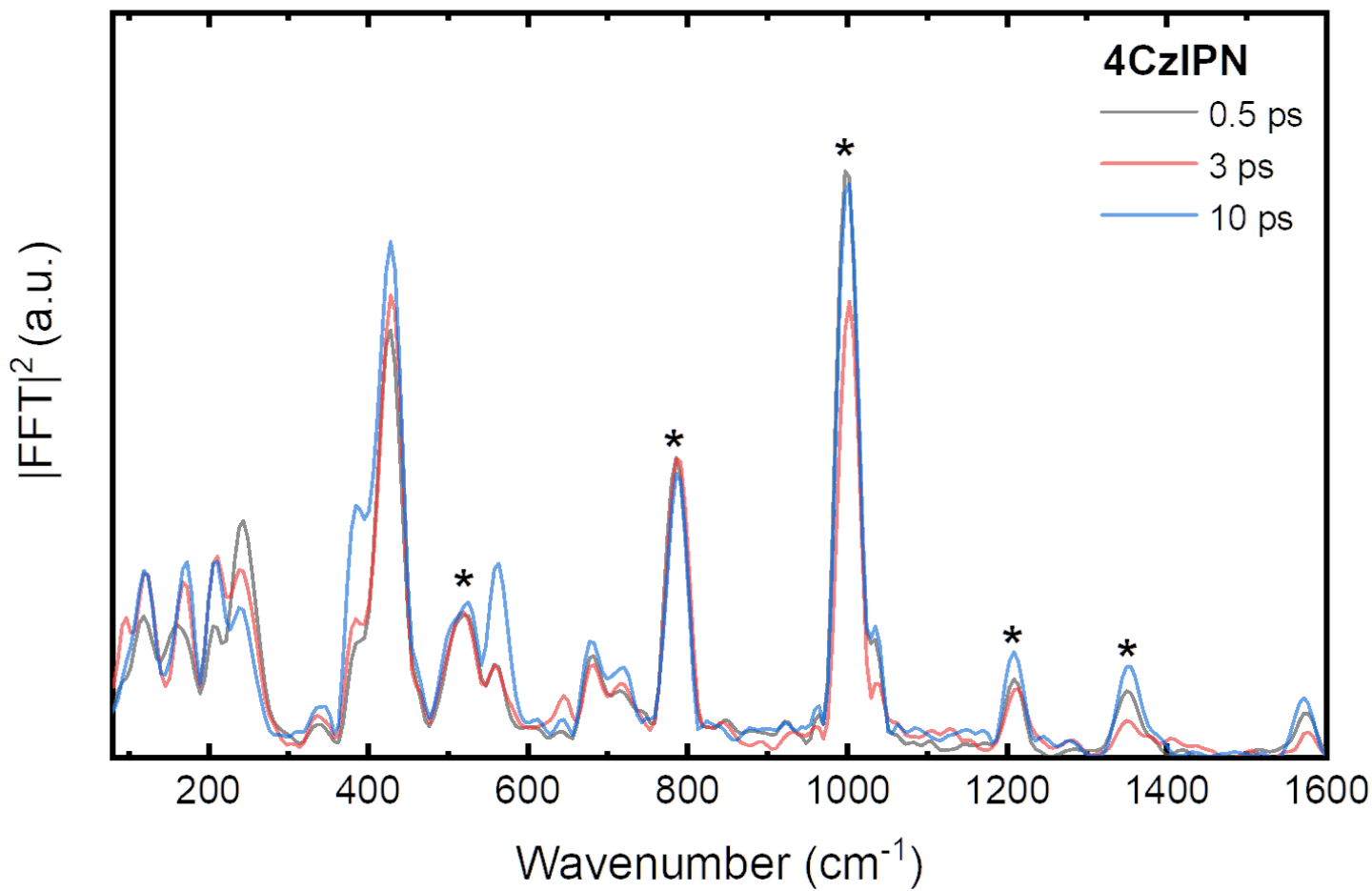

**Figure S20:** The IVS spectra of 4CzIPN in toluene taken at 0.5, 3 and 10 ps after excitation to isolate the modes that are associated with the $^1$CT excited state of 4CzIPN (toluene solvent modes are marked by black asterisks). The IVS was obtained by first exciting 4CzIPN at 450 nm with a 200 fs pulse (fluence 0.47 mJ/cm$^2$), before a second pulse at 850 nm with a 8.5 fs duration was used (5.7 mJ/cm$^2$) to induce coherent oscillations in the excited state of 4CzIPN. As time progresses, a mode at 239 cm$^{-1}$ decreases in intensity, whilst modes at 114, 168, 206, 385, 429, 565, 684, 717 and 1574 cm$^{-1}$ increase in intensity.

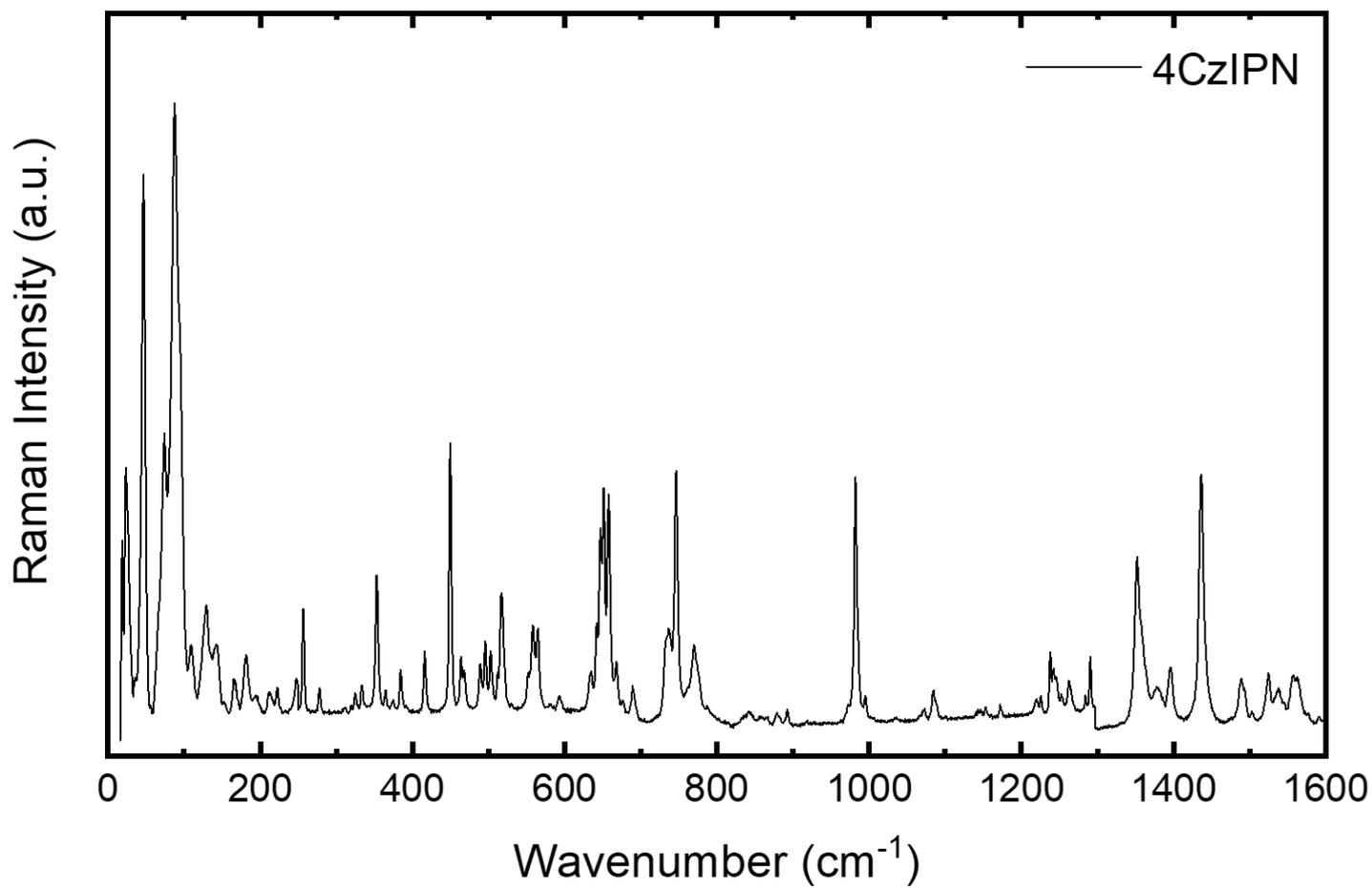

**Figure S21:** The off-resonant (633 nm) Raman spectra of a neat 4CzIPN film.

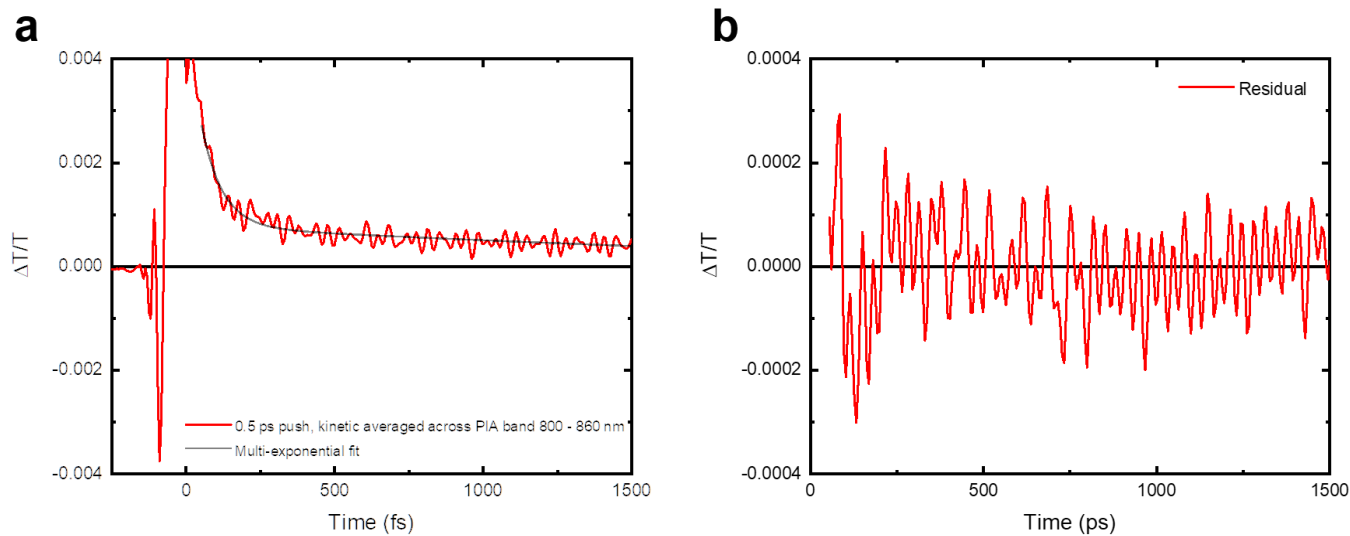

**Figure S22: (a)** The pump-push-probe kinetic taken from the $^1$CT PIA (800-860 nm) of 4CzIPN in a toluene solution, excited with a 200 fs 450 nm pulse and pushed 500 fs after the pump pulse with a 8.5 fs pulse centered at 850 nm, chosen to be resonant with the 4CzIPN $^1$CT PIA. The kinetic was fitted with a multi-exponential function to isolate the oscillations induced in the excited state by the fast push pulse. **(b)** The residual oscillations in the $^1$CT excited state extracted from the pump-push-probe kinetics. The vibrational modes associated with these oscillations were then extracted using a FFT.

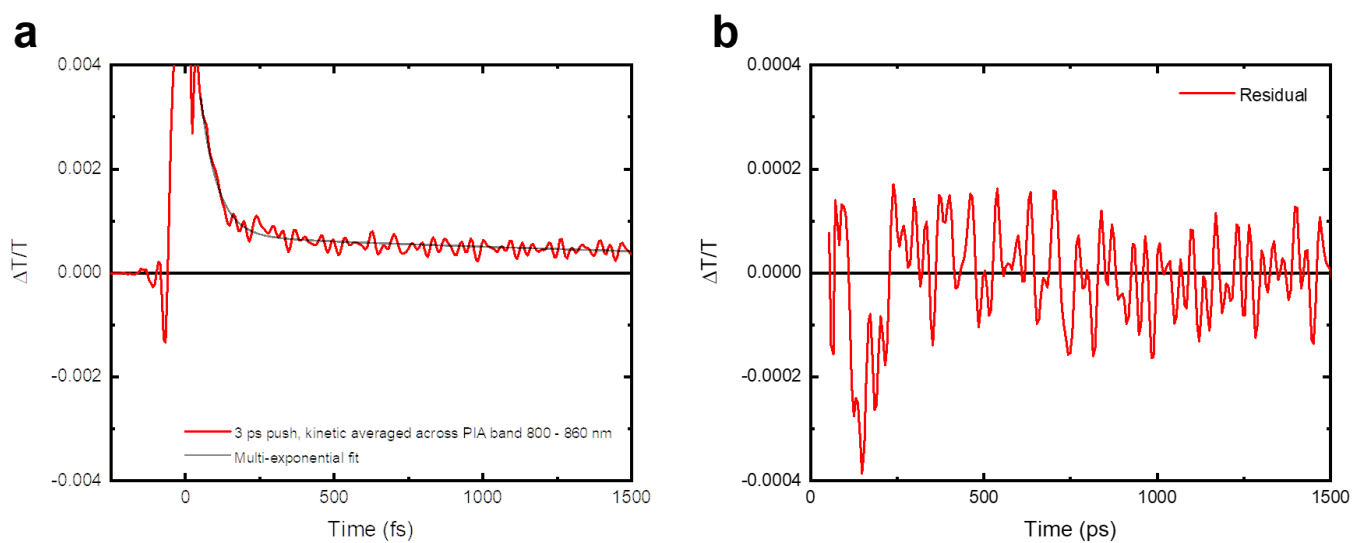

**Figure S23: (a)** The pump-push-probe kinetic taken from the $^1$CT PIA (800-860 nm) of 4CzIPN in a toluene solution, excited with a 200 fs 450 nm pulse and pushed 3 ps after the pump pulse with a 8.5 fs pulse centered at 850 nm, chosen to be resonant with the 4CzIPN $^1$CT PIA. The kinetic was fitted with a multi-exponential function to isolate the oscillations induced in the excited state by the fast push pulse. **(b)** The residual oscillations in the $^1$CT excited state extracted from the pump-push-probe kinetics. The vibrational modes associated with these oscillations were then extracted using a FFT.

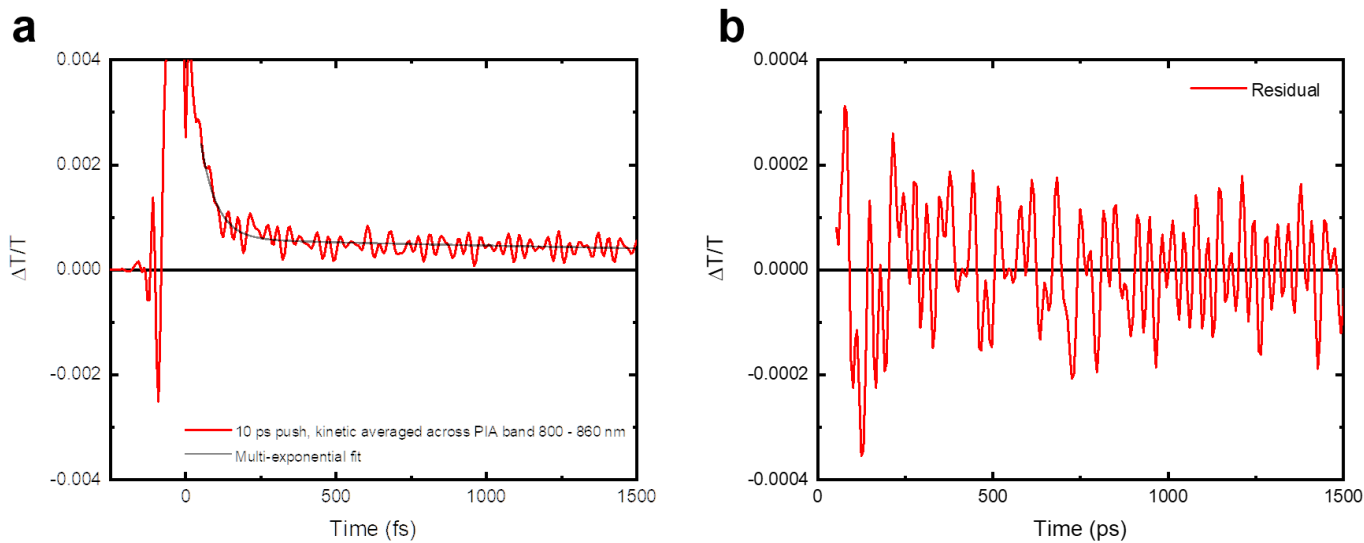

**Figure S24:** (a) The pump-push-probe kinetic taken from the $^1$CT PIA (800-860 nm) of 4CzIPN in a toluene solution, excited with a 200 fs 450 nm pulse and pushed 10 ps after the pump pulse with a 8.5 fs pulse centered at 850 nm, chosen to be resonant with the 4CzIPN $^1$CT PIA. The kinetic was fitted with a multi-exponential function to isolate the oscillations induced in the excited state by the fast push pulse. (b) The residual oscillations in the $^1$CT excited state extracted from the pump-push-probe kinetics. The vibrational modes associated with these oscillations were then extracted using a FFT.

# Computational details

In the calculations, we focused on the time-evolution of the singlet- and triplet-states of TXO-TPA in a dilute solution utilizing a combined quantum mechanics/molecular mechanics (QM/MM) approach. The QM part contained one TXO-TPA molecule, described at the density functional theory (DFT) level using the PBEh-3c functional[1], while the surrounding 249 toluene molecules were treated with the GROMOS 54a7 classical force field, as derived by ATB[2]. We used an electrostatic embedding scheme, that is interaction with the solvent includes electronic polarization of the QM TXO-TPA molecule by partial charges on the MM toluene molecules, together with Lennard-Jones potentials. With this setup, we first equilibrated the system at 300 K and 0.1 MPa, as controlled by the Nose-Hoover thermostat[3] and Parrinello-Rahman barostat[4], and achieved the resulting density of 0.85 g/cm$^3$. Next, we performed adiabatic excited-state dynamics in the lowest singlet $S_1$ excited state (aka $^1$CT) for 10 ps using the forces from time-dependent (TD) DFT. The same approach was further utilized to study the triplet dynamics, yet now sampling the geometries from 10 ps of adiabatic dynamics in the lowest triplet state from open-shell DFT calculations, before applying spin-adapted TD-DFT calculations to access the properties of higher-lying triplet excitons. The QM/MM calculations were performed via an inbuild interface between the ORCA[5] and Gromacs[6] softwares. For the analysis of charge-transfer properties, we used the Multiwfn code[7]. Benchmark calculations of the excitation energies were also performed by SCS-CC2 method[8] using cc-pVDZ basis set[9] as implemented in TURBOMOLE 6.5[10].

The solvent dipole moment was evaluated based on the difference in molecular dipole moments, which arises due to the point charges and is given as follows:

$$d_{solv} = |\vec{d}_{with\ PC} - \vec{d}_{w/o\ PC}|$$

where $d_{with\ PC}$ and $d_{w/o\ PC}$ are molecular dipoles at a given geometry in the presence of point charges and in vacuum, respectively. $d_{with\ PC}$ and $d_{w/o\ PC}$ were determined along both singlet- and triplet-dynamics for a given active state (either $S_1$ or $T_1$), so that the resulting solvent dipole moments also incorporate a contribution from the change in the state character (from CT to LE or vice versa) due to the polarisation, as observed along the triplet dynamics.

The calculations of the (time-dependent) geometrical changes were based on the ground state Hessian matrix as implemented in ORCA. Here, at each time step, the structure was represented by a set of normal coordinates, enabling us to track the difference in the displacements relative to a reference geometry. Note that in our calculations we focus on the modes with the frequencies below 1000 cm$^{-1}$, regardless of their Raman activity. In turn, the intensity of a mode was associated with a value of standard deviation, which was computed from the fluctuations upon dynamics within the indicated time intervals.

|   | Singlets | | Triplets | |
|---|---|---|---|---|
|   | SCS-CC2 | PBEh-3c | SCS-CC2 | PBEh-3c |
| 1 | **3.6** | **3.3** | **3.2** | **2.9** |
| 2 | 3.8 | 3.8 | **3.5** | **3.2** |
| 3 | 4.1 | 4.1 | **3.7** | **3.4** |

**Table S1:** Vertical excitation energies form the ground state (PBEh-3c) geometry, obtained with SCS-CC2 and PBEh-3c. For the target states, marked by bold, both methods show good agreement, while the PBEh-3c excitation energies are consistently lower by 0.3 eV. Triplet 1 is the $^3$CT state, whilst triplet 2 is a $^3$LE state localized on the TXO moiety.

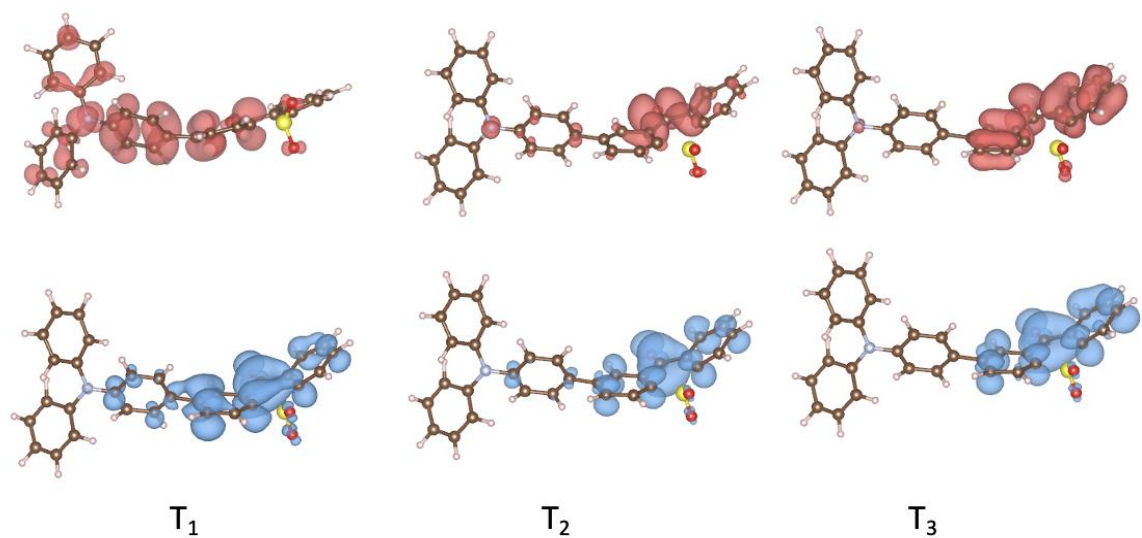

$T_1$          $T_2$          $T_3$

**Figure S25:** Electron-hole densities computed for the three lowest triplet states at the ground state geometry, showing that $T_1$ has mixed CT-LE character, while $T_2$ and $T_3$ are LEs located on the TXO (acceptor) part of TXO- TPA.

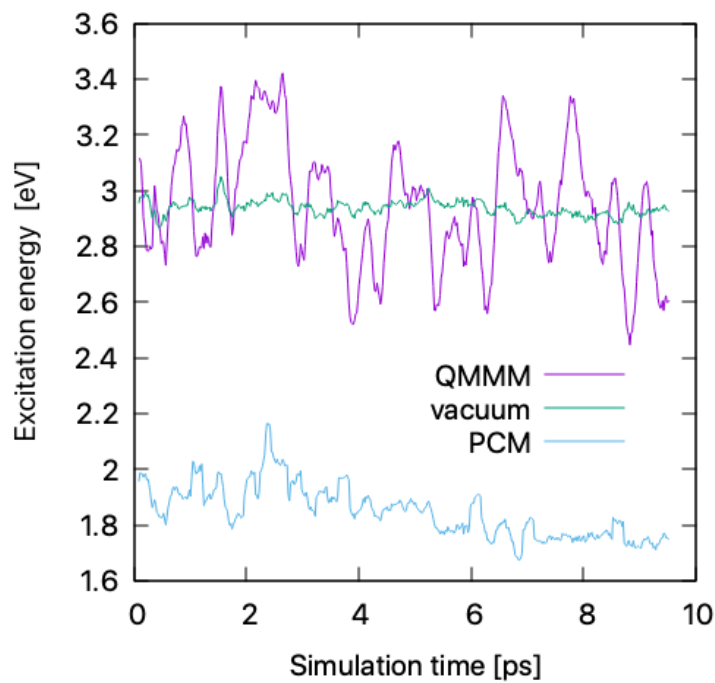

**Figure S26:** Modulation of the $S_1$ excitation energy along the adiabatic dynamics caused by the different treatment of the environment. Here, at each time step, the geometry of the emitter was extracted from the $S_1$ trajectory, while solvent is taken into account either explicitly (QM/MM) or by means of PCM or is completely absent (vacuum).

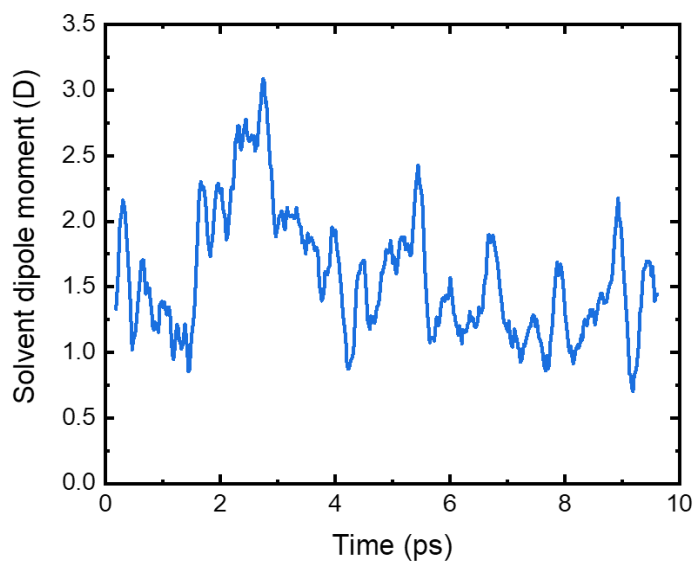

**Figure S27:** The solvent dipole moment over the $^1$CT simulation trajectory. The solvent dipole moment peaks at around 2.5 ps, reflecting the highly polarised environment that is shown in the 2.5 ps image of Fig. 4c.

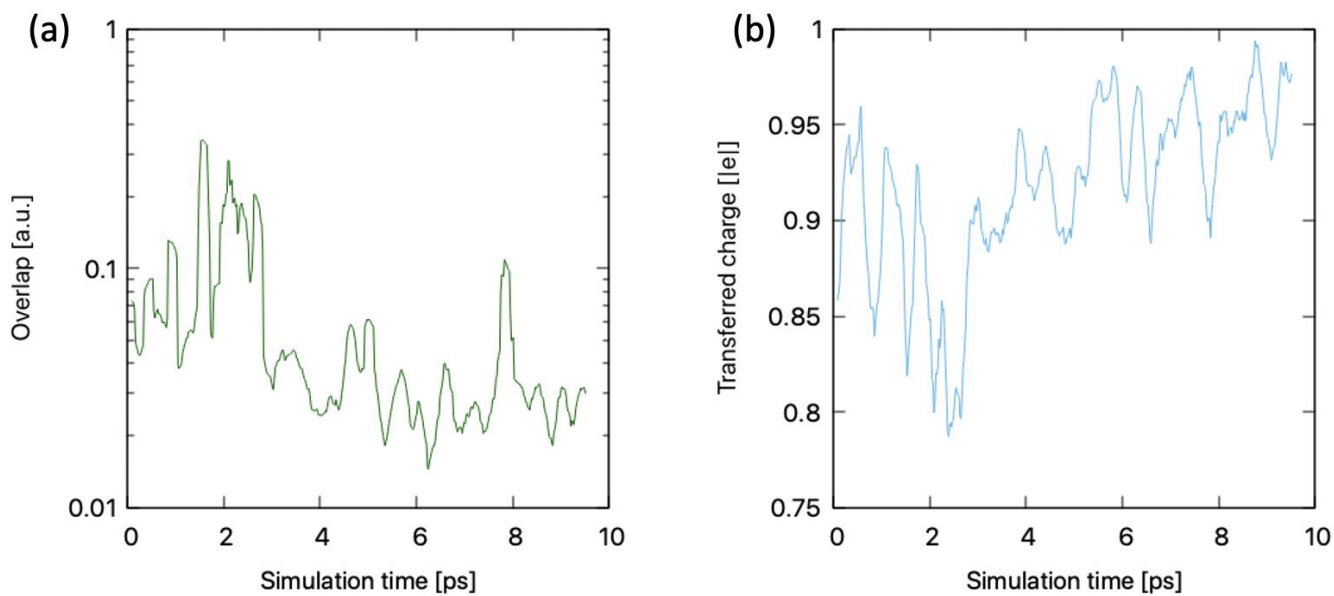

**Figure S28:** Time evolution of **(a)** electron-hole density overlap and **(b)** transferred charge computed for the $S_1$ state along the adiabatic dynamics.

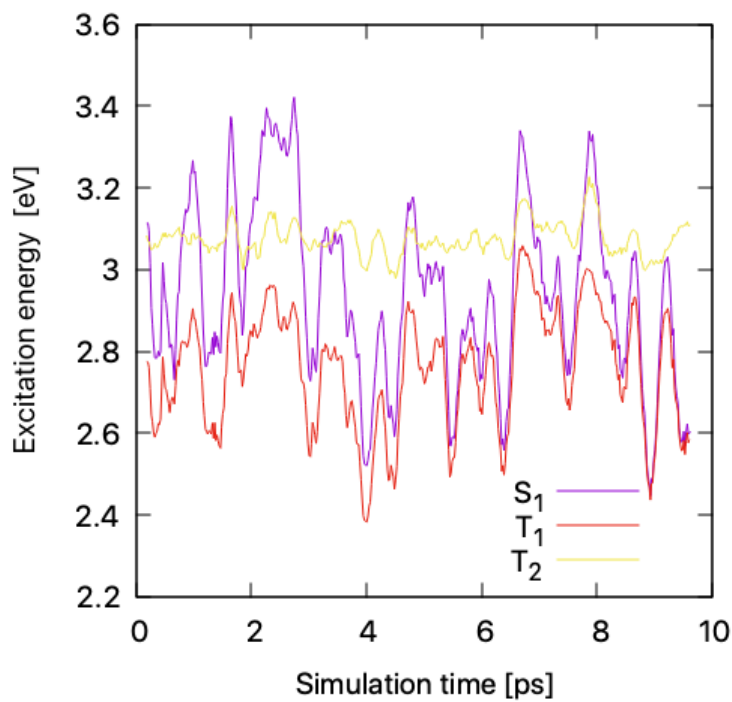

**Figure S29:** Time evolution of the excited state energies from the $S_1$ adiabatic dynamics. Here, the CT-like $T_1$ is strongly affected by the solvent polarization, while the LE-like $T_2$ essentially oscillates around the mean value.

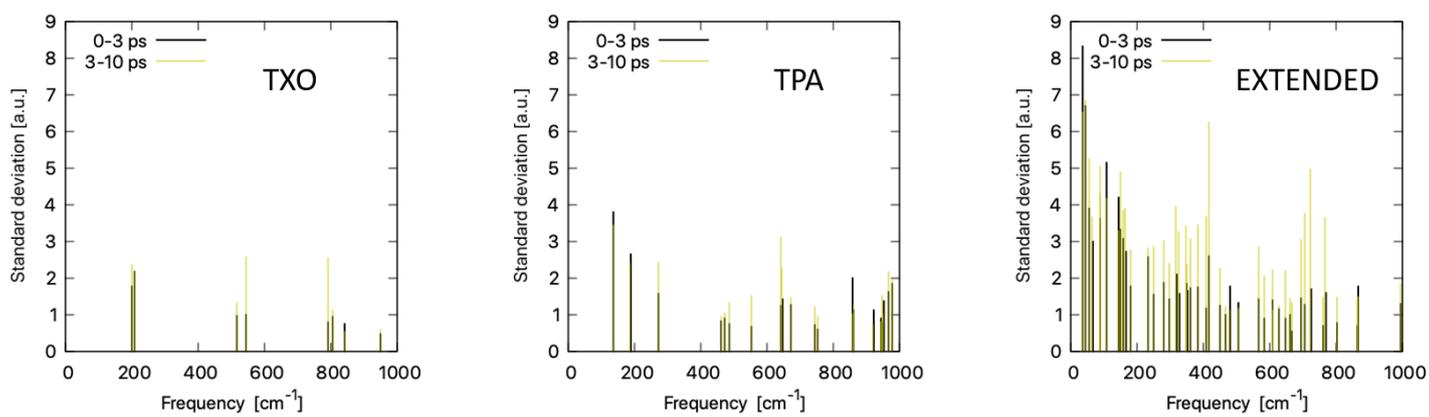

**Figure S30:** The time evolution and localization of the excited state vibrational modes of TXO-TPA, taken from the $S_1$ adiabatic dynamics. Modes localized on the D (TPA) or the A (TXO) are defined as those with a contribution of <15% from the A or D, respectively.

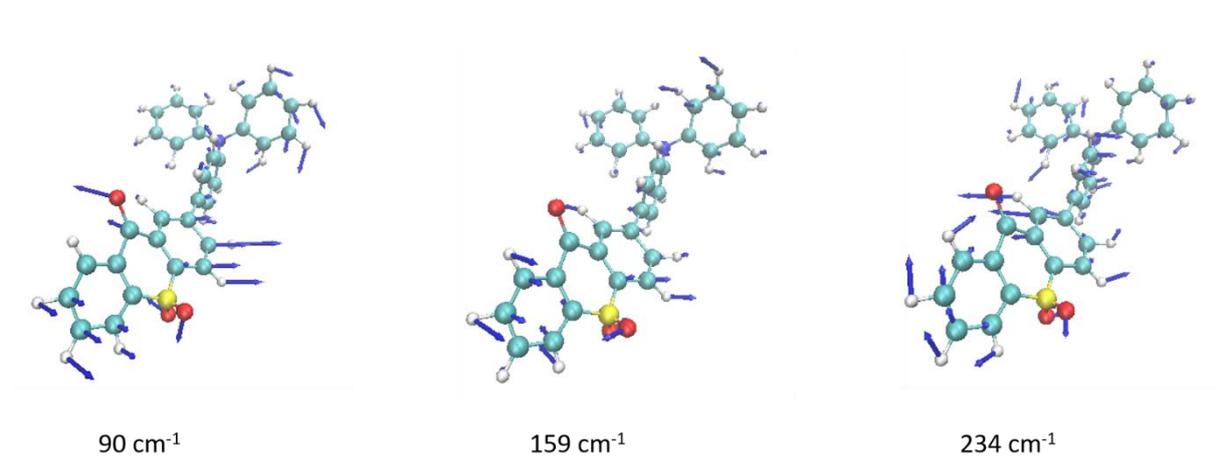

90 cm⁻¹  159 cm⁻¹  234 cm⁻¹

**Figure S31:** The low frequency spectator modes present in the singlet excited state manifold of TXO-TPA.

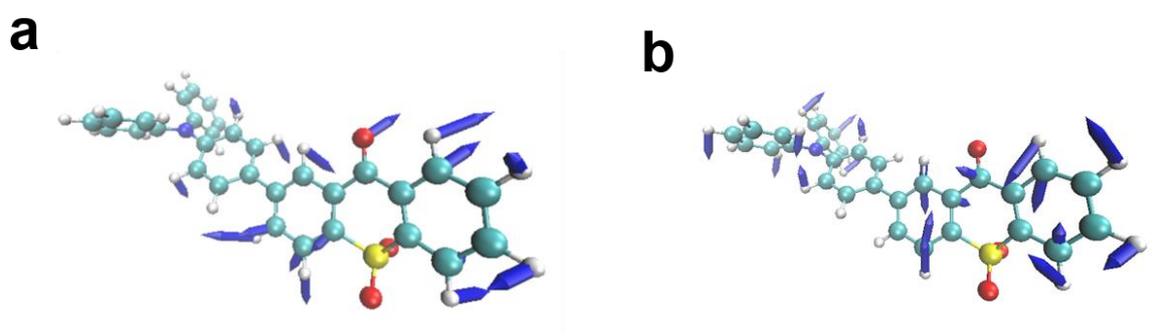

**Figure S32: (a)** The 417 cm$^{-1}$ mode present after formation of the quasi-pure $^1$CT state in TXO-TPA. **(b)** The 712 cm$^{-1}$ mode present after formation of the quasi-pure $^1$CT state in TXO-TPA.

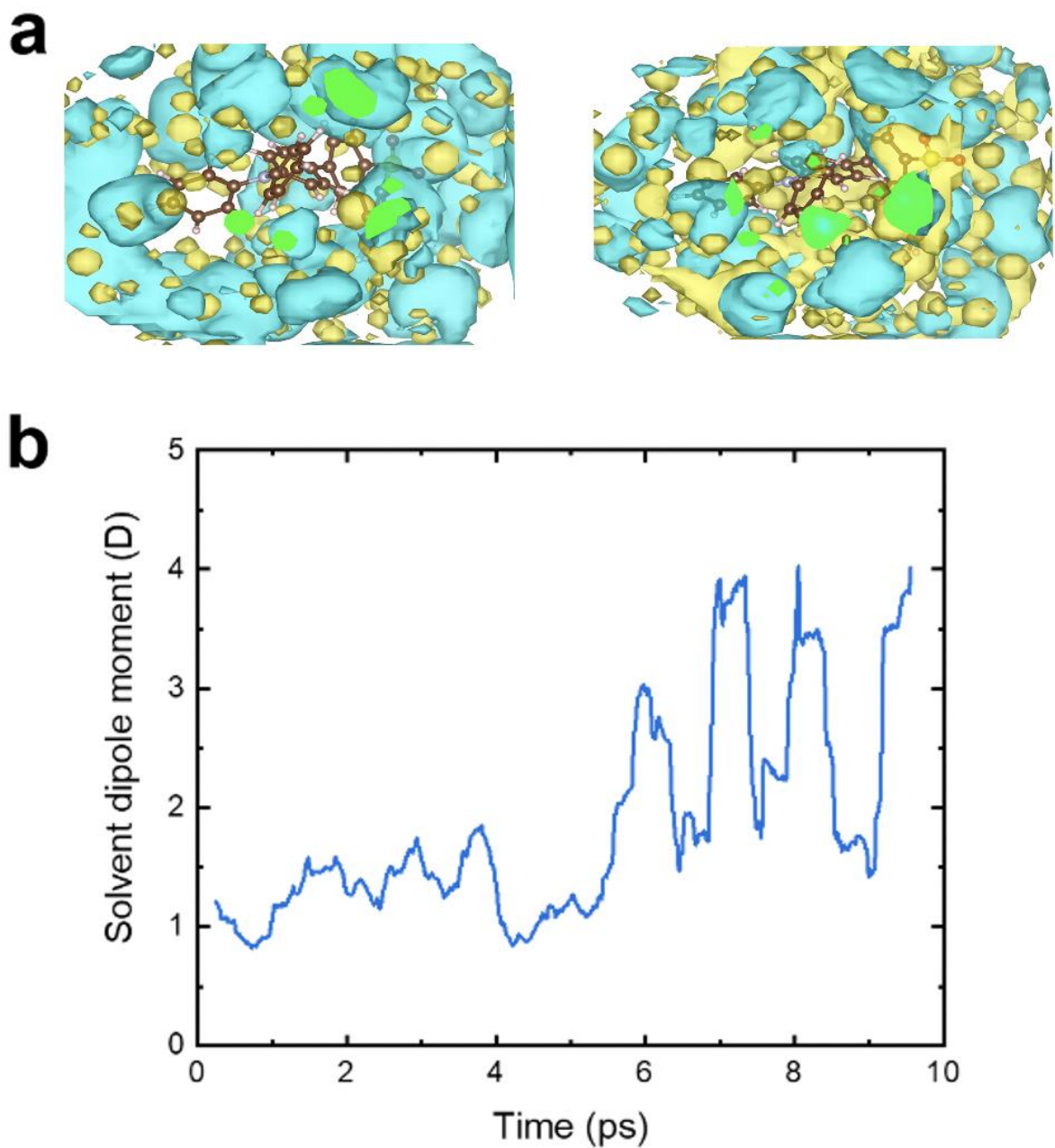

**Figure S33: (a)** The electrostatic potential from the solvent at 5 ps (left) and 6 ps (right) in the simulations of the triplet manifold. **(b)** The solvent dipole moment over the simulation trajectory. The solvent dipole moment increases rapidly >5 ps when the more polar 3CT state becomes the lowest energy, and therefore populated, triplet state.

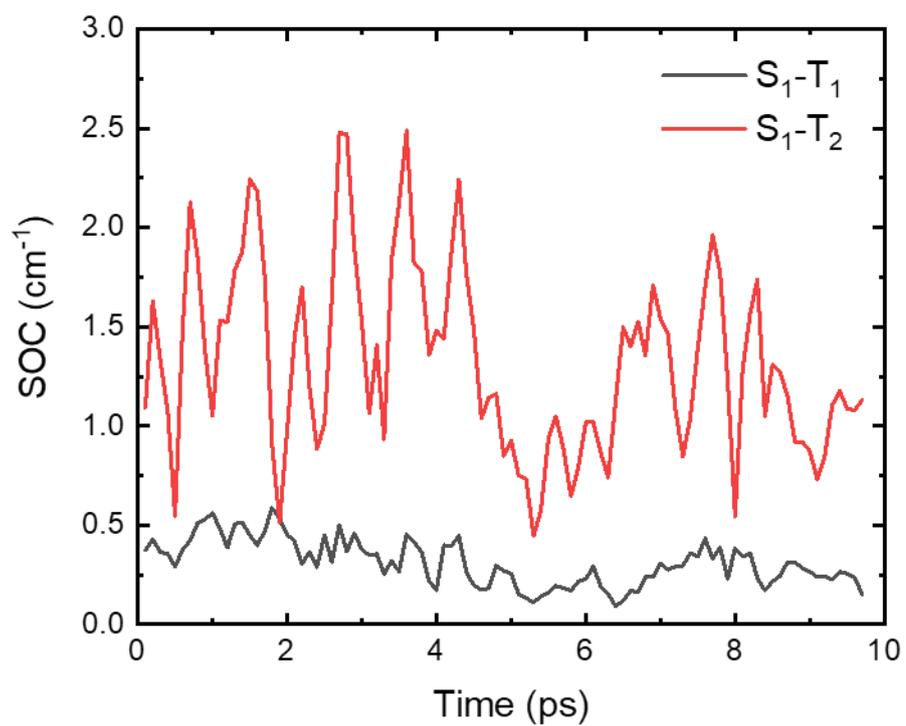

**Figure S34:** The SOC matrix elements computed between $S_1$-$T_1$ and $S_1$-$T_2$ along the triplet simulation trajectory.

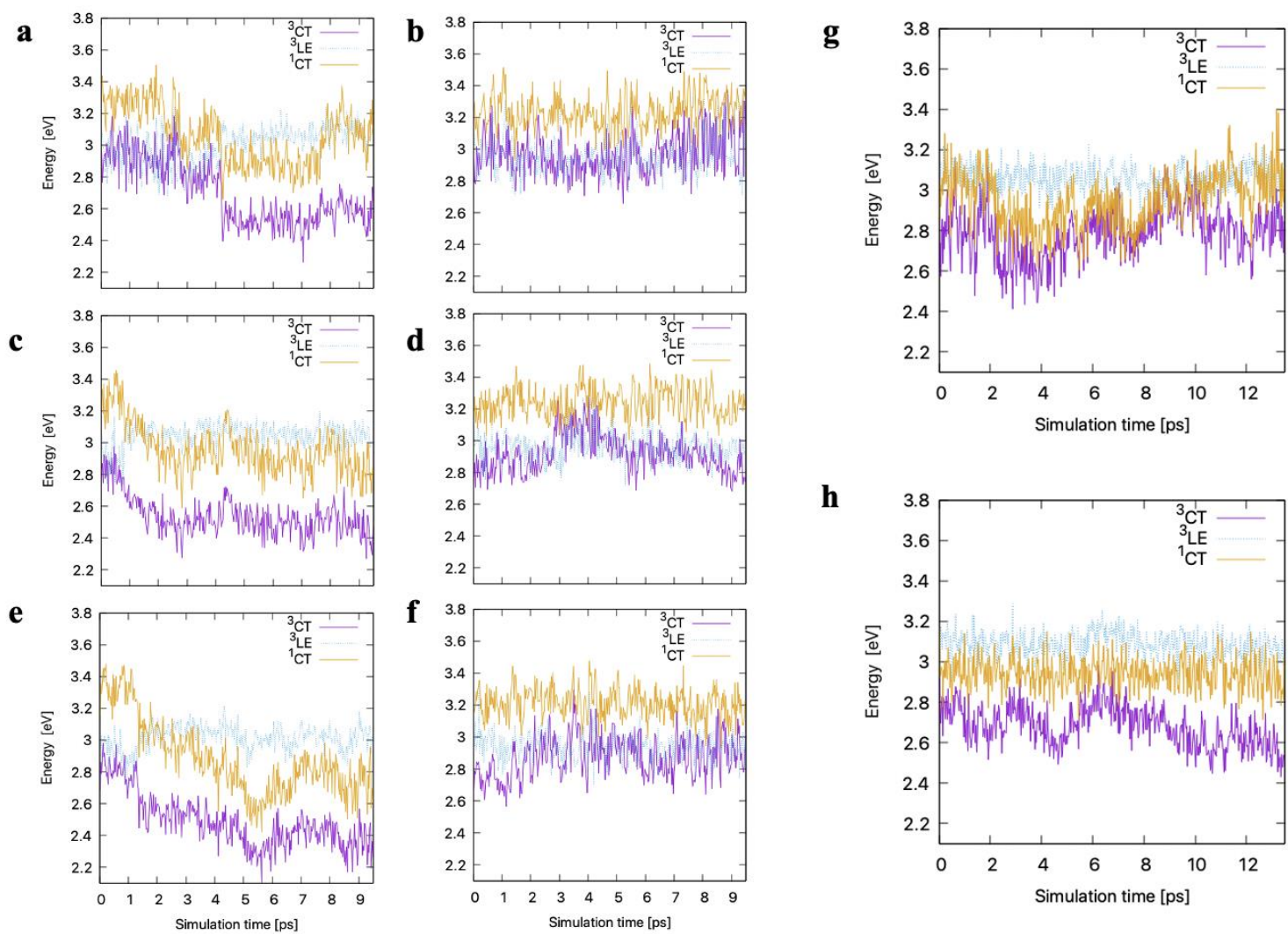

**Figure S35: (a-f)** Triplet dynamics of TXO-TPA, starting from different initial configurations of the emitter. The trajectories **a**, **c**, and **e** were obtained in the presence of polarizable solvent using the QM/MM methodology, while **b**, **d**, and **f** were calculated in vacuum. The trajectories **a** and **b**, **c** and **d**, **e** and **f** share the same initial configurations of TXO-TPA, respectively. The initial conditions were extracted from 10 ps of ground state dynamics in toluene. (**g,h**) The dynamics of TXO-TPA in the lowest (excited) singlet state, as computed (**g**) in toluene and (**h**) in vacuum.

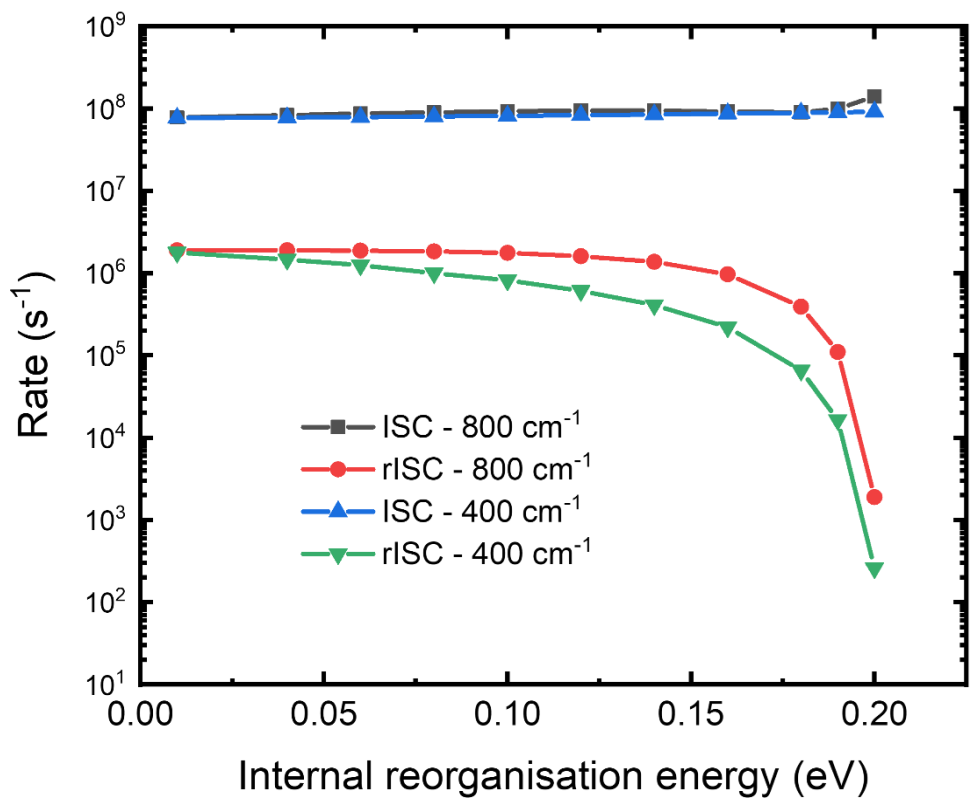

**Figure S36:** ISC and rISC rates for TXO-TPA, calculated using semi-classical Marcus-Levich-Jortner theory. Here, the vibrational modes treated quantum mechanically are the 400 and 800 cm$^{-1}$ modes identified in our work. The internal reorganization energy is varied whilst the total reorganization energy is kept constant. Little influence is found for the rate of ISC, as population transfer from S$_1$ to T$_1$ can still occur via tunneling to the discrete vibronic replica. However, rISC is slowed considerably due to the absence of an isoenergetic S$_1$ vibronic sublevel for tunneling to occur to from T$_1$ (E$_{S1}$ > E$_{T1}$). We note that an excellent agreement to the experimentally observed k$_{rISC}$ of 2.3×10$^5$ s$^{-1}$ is found for an internal reorganisation energy contribution of ~0.16-0.19 eV.